\documentclass[sigconf]{acmart}

\AtBeginDocument{%
  \providecommand\BibTeX{{%
    \normalfont B\kern-0.5em{\scshape i\kern-0.25em b}\kern-0.8em\TeX}}}

\setcopyright{rightsretained}
\copyrightyear{2021}
\acmYear{2021}
\acmConference[CHI '21]{CHI Conference on Human Factors in Computing Systems}{May 8--13, 2021}{Yokohama, Japan}
\acmBooktitle{CHI Conference on Human Factors in Computing Systems (CHI '21), May 8--13, 2021, Yokohama, Japan}\acmDOI{10.1145/3411764.3445443}
\acmISBN{978-1-4503-8096-6/21/05}

\usepackage{subfigure,multirow}

\definecolor{purple}{HTML}{7d41a8}
\definecolor{yellow}{HTML}{fbcb28}
\definecolor{fuchsia}{HTML}{ff3ac9}
\definecolor{blue}{HTML}{91c0d3}
\definecolor{brown}{HTML}{ab762b}

\newcommand{\purple}[1]{\textbf{\textcolor{purple}{#1}}}

\newcommand{\fuchsia}[1]{\textbf{\textcolor{fuchsia}{#1}}}
\newcommand{\blue}[1]{\textbf{\textcolor{cyan}{#1}}}
\newcommand{\brown}[1]{\textbf{\textcolor{brown}{#1}}}

\begin{document}

\title{\editmark{Towards Understanding How Readers Integrate Charts and Captions: A Case Study with Line Charts}}

\author{Dae Hyun Kim}
\email{dhkim16@cs.stanford.edu}
\affiliation{
	\institution{Stanford University/Tableau Research}
	\city{Stanford/Palo Alto}
	\state{California}
	\postcode{94305, 94306}
}

\author{Vidya Setlur}
\email{vsetlur@tableau.com}
\affiliation{
	\institution{Tableau Research}
	\city{Palo Alto}
	\state{California}
	\postcode{94306}
}

\author{Maneesh Agrawala}
\email{maneesh@cs.stanford.edu}
\affiliation{
	\institution{Stanford University}
	\city{Stanford}
	\state{California}
	\postcode{94305}
}


\definecolor{darkgreen}{rgb}{0,0.5,0}
\definecolor{orange}{rgb}{1,0.5,0}
\definecolor{teal}{rgb}{0,0.5,0.5}
\definecolor{darkpurple}{rgb}{0.5, 0, 0.5}

\newcommand {\daehyun}[1]{{\color{blue}\bf{DH: #1}\normalfont}}
\newcommand {\vidya}[1]{{\color{teal}\bf{VS: #1}\normalfont}}
\newcommand {\maneesh}[1]{{\color{red}\bf{MA: #1}\normalfont}}
\newcommand {\change}[1]{{\color{orange}#1\normalfont}}
\newcommand{\cut}[1]{}



\newcommand{\singlecol}[1]{}
\newcommand{\doublecol}[1]{#1}


\newcommand{\editmark}[1]{#1}
\newcommand{\editmarkvar}[1]{#1}

\newenvironment{tight_itemize}{\begin{itemize} \itemsep
-2.1pt}{\end{itemize}}

\newenvironment{tight_enumerate}{\begin{enumerate} \itemsep
-2.1pt}{\end{enumerate}}

\begin{abstract}
Charts often contain visually  \editmark{prominent} features that draw attention to aspects of the data and include text captions that emphasize aspects of the data. Through a crowdsourced study, we explore how readers gather takeaways when considering charts and captions together. We first ask participants to mark visually  \editmark{prominent} regions in a set of line charts. We then generate text captions based on the  \editmark{prominent} features and ask participants to report their takeaways after observing chart-caption pairs. We find that when both the chart and caption describe a  \editmark{high-prominence} feature, readers treat the doubly emphasized  \editmark{high-prominence} feature as the takeaway; when the caption describes a  \editmark{low-prominence} chart feature, readers rely on the chart and report a  \editmark{higher-prominence} feature as the takeaway. We also find that external information that provides context, helps further convey the caption’s message to the reader. We use these findings to provide guidelines for authoring effective chart-caption pairs.
\end{abstract}



\singlecol{
\begin{teaserfigure}
\centering
  \includegraphics[width=.7\columnwidth]{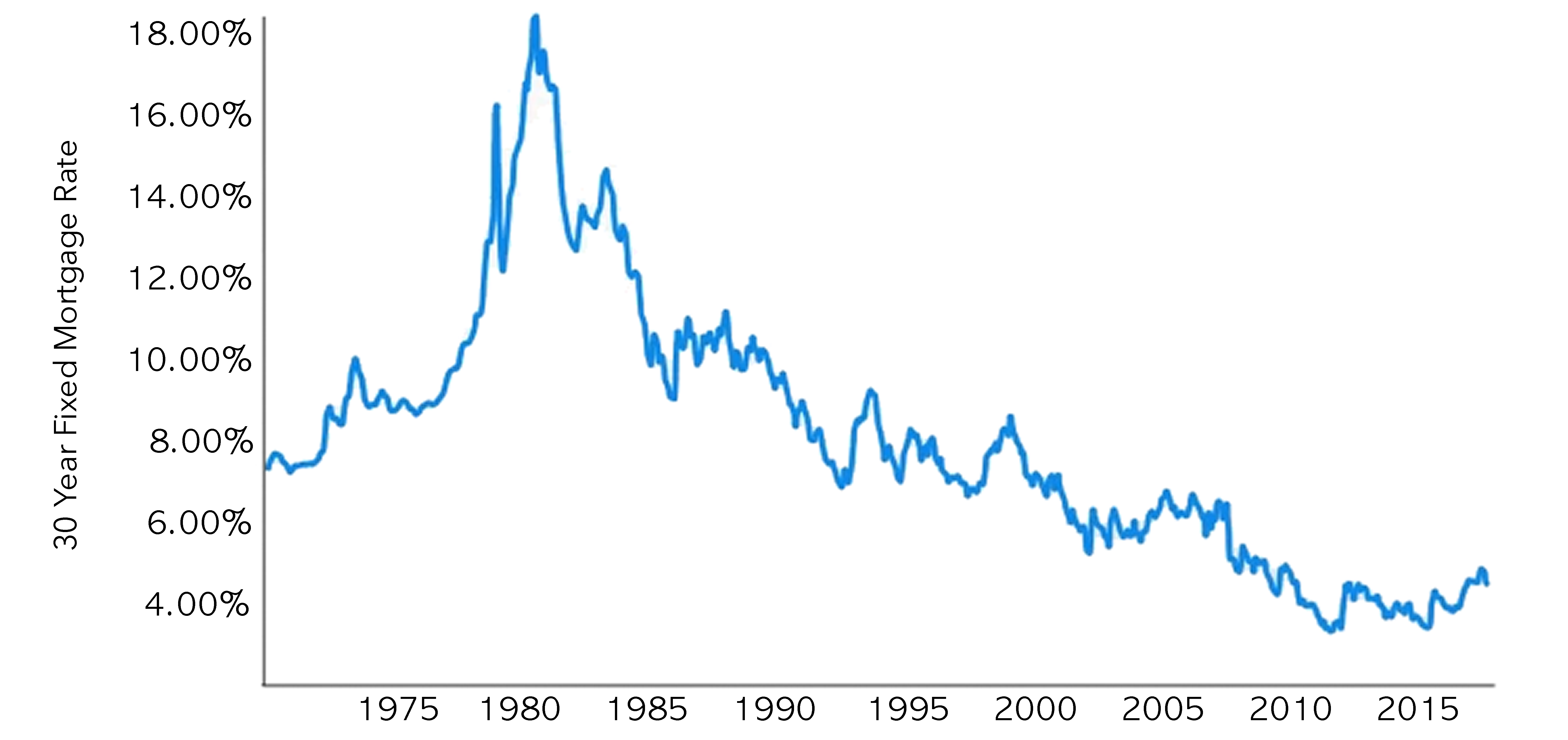}
\caption*{}
  \label{fig:teaser}
  \Description[An example line graph]{An example line graph showing 30 year fixed mortgage rates.}
  \vspace{-4mm}
\end{teaserfigure}
}

\maketitle


\section{Introduction}
Charts provide graphical representations of data that can draw a
reader's attention to various visual features such as outliers and
trends. \editmark{Readers are initially drawn towards the most \textit{visually salient} components in the chart such as the chart title and the labels~\cite{matzen2017data}. However, they eventually apply their cognitive processes to extract meaning from the most \textit{prominent} chart
features}~\cite{card:1999,tufte:1990}. 
Consider the line
chart at the beginning of this article. What do you think are the main
visual features of the chart and what are its key takeaways?

\editmark{Such charts are often accompanied by text captions that emphasize specific aspects of the data as chosen by the chart author. In some cases, the data emphasized in the caption corresponds to the most
visually prominent features of the chart and in other cases it does
not. Prior studies have shown that charts with captions can
improve both recall and comprehension of some aspects of
the underlying information, compared to seeing the chart or the
caption text
alone~\cite{bransford1979human,nugent:1983,large:1995,hegarty:1993}.
But far less is known about how readers integrate information between
charts and captions, especially when the data emphasized by the
visually \editmark{prominent} features of the chart differs from the data that is
emphasized in the caption.}

\doublecol{
\begin{figure}
\centering
  \includegraphics[width=\linewidth]{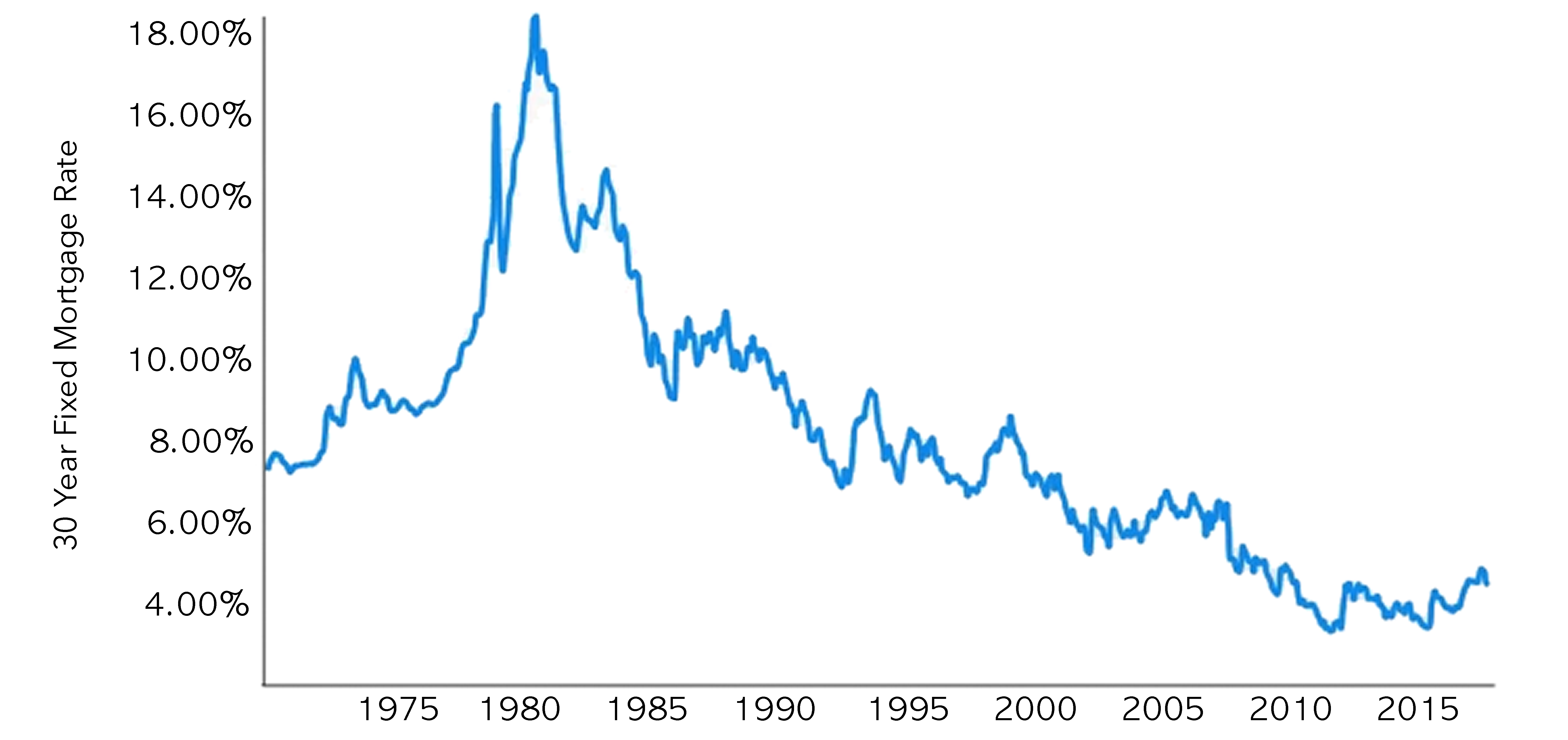}
  \label{fig:teaser}
  \Description[An example line graph]{An example line graph showing 30 year fixed mortgage rates.}
  \vspace{-10mm}
\end{figure}
}

\editmark{Consider the visually prominent features in our initial line chart and
then consider each of the following caption possibilities one at a
time. How do your takeaways change with each one?}\\[3pt]
{\small
{\bf \em (1)} The chart shows the 30-year fixed mortgage rate between 1970 and 2018.\\[1pt]
{\bf \em (2)} The 30-year fixed mortgage rate increased slightly from 1997 to 1999.\\[1pt]
{\bf \em (3)} The 30-year fixed mortgage rate reached its peak of 18.45\% in 1981.\\[1pt]
{\bf \em (4)} The 30-year fixed mortgage rate reached its peak of 18.45\% in 1981 due to runaway inflation.\\[3pt]
  }
\editmark{The first caption simply describes the dimensions graphed in the chart
and only provides redundant information that could be read from the
axis labels. Automated caption generation tools often
create such {\em basic} descriptive
captions\,\cite{tableau,powerbi}. The next three captions each
emphasizes aspects of the data corresponding to a visual feature of the
chart (i.e., upward trend, peak) by explicitly mentioning the corresponding data point or trend.
However, the second caption emphasizes a feature of low visual prominence -- a relatively local and small rise in the chart between
1997 and 1999. The third caption describes the most visually prominent
feature of the chart -- the tallest peak that occurs in 1981. The
fourth caption also describes this most visually prominent feature, but
adds external information that is not present in the chart and
provides context for the data.}

In this paper, we examine two main hypotheses - (1) When a caption
emphasizes more visually prominent features of the chart, people are
more likely to treat those features as the takeaway\editmark{; even when a
  caption emphasizes a less visually prominent feature, people are still more
  likely to treat a more visually prominent feature in the chart as the
  takeaway}. (2) When a caption contains external information for
context, the information serves to further emphasize the feature
described in the caption and readers are therefore more likely to
treat that feature as the takeaway.

\editmark{We considered univariate line charts for our work because they are among the most common basic charts and are easily parameterizable, making them useful for the initial exploration of our hypotheses. We synthesized $27$ single charts with carefully chosen parameters and collected $16$ real-world single line charts to confirm the generalizability of our findings.} \editmark{We ran a data collection activity} on \editmark{the} $43$ \editmark{single-}line charts, where we asked \editmark{$219$} participants
to mark visually prominent regions on the line charts. We
generated text captions for the ranked set of prominent features
\editmark{using templates to control variations in natural language. Finally, we} conducted a crowdsourced study with a new set of \editmark{$2168$}
participants to report their takeaways after seeing the chart-caption
pairs.

\editmark{Our findings from the study support both of our hypotheses. Referring back to our initial line
chart, when the caption mentions the most prominent feature as in the third caption (i.e., the peak in 1981), readers will probably take away information from that feature. When the caption mentions a less prominent feature as in the second caption (i.e., the increase from 1997 to 1999), there is a mismatch in the message between the chart and the caption. Readers will have a strong tendency to go with the message conveyed in the chart and take away information about the peak value. Finally, the external information about the peak value present in the fourth caption will reinforce the message in the caption and the readers will more likely take away information about the peak.}

These findings help better understand the relationship between charts
and their captions when conveying information about certain aspects of
the data to the reader. Based on these studies, we provide guidelines
for authoring charts and captions together in order to emphasize the
author's intended takeaways. \editmark{Visualization authors can more effectively convey their message to readers by ensuring that both charts and captions emphasize the same set of features.}
Specifically, \editmark{authors could make visual features that are related to their key message, more prominent} through visual cues (e.g., highlighting or zooming into a focus area, adding annotations)~\cite{Egeth1997VisualAC,5571352} \editmark{or include external information in the caption to further emphasize the feature described in the caption}. Often, an alternative chart representation may be more conducive to making certain visual
features more prominent.

\section{Related Work}

\editmark{Our work is related to two lines of research: (1) Cognitive Understanding of Charts and (2)  Caption Generation Tools.}

\subsection{Cognitive Understanding of Charts}
The prevalence of text with visuals has led researchers to explore how readers
specifically understand information in figures with accompanying text in several
domains. Li et al.~\cite{li:2018} conducted studies to demonstrate
that figures with text can convey essential information and better aid
understanding than just text alone for scientific publications in a
biomedical domain. Odell et al.~\cite{ottley:2016} demonstrated that
having text that accurately describes important findings in medical
diagnostic images can increase physicians' speed and accuracy on
Bayesian reasoning tasks while making life-critical judgments for
patients. Xiong et al.~\cite{xiong2019curse} showed that background
knowledge can affect viewers' visual perception of data as they tend
to see the pattern in the data corresponding to the background
knowledge as more visually salient.  Kong et
al.~\cite{Kong2018FramesAS} explored the impact of titles on
visualization interpretation \editmark{with different degrees of misalignment between a visualization and its title.}  A title contains a miscued slant when the visualization emphasizes
one side of the story through visual cues but the title’s message addresses the other (less emphasized) side of the story. Titles have a contradictory slant where the information conveyed in the
title is not presented at all in the visualization. They observe that even though the title of a visualization may not be recalled, the title can still measurably impact the remembered contents of a chart. Specifically, titles with a contradictory slant trigger more people to identify bias compared to titles with a miscued slant, while visualizations are perceived as impartial by the
majority of viewers~\cite{kong:2019}. Elzer et al.~\cite{elzer2005exploring} conducted a study to better
understand the extent to which captions contribute to recognizing the
intended message of an information graphic for sight-impaired
users. They find that the caption strengthens
the intended message of the graphic.
\editmark{Carberry et al.~\cite{carberry:2006} showed that the
  communicative goals of infographics in digital libraries are often
  not repeated in the text of the articles. Their work looked into how information in the graphics could be better utilized for
  summarizing a document by employing a Bayesian network.}   

\editmark{However, this previous research has not explored the relationship between charts and their captions with respect to how they work together to emphasize certain aspects of the data to the reader.}

\subsection{Caption Generation Tools}
\editmark{ A number of visual analysis tools help users design charts and captions from an input data table~\cite{datasite,demiralp2017foresight,hu2018dive,Vartak2015,wills2010autovis}. These captions  generally only describe the data attributes and visual encodings that are in play in the charts and do not highlight key takeaways.} \editmark{Nevertheless, authors often include text with a chart to
  help emphasize an intended message to their audience.} PostGraphe~\cite{fasciano1996postgraphe} generated
reports integrating graphics and text from a list of user-defined
intentions about the underlying data such as the comparison of
variables. SAGE~\cite{mittal:1995} used natural language generation
techniques to produce explanatory captions for information
graphics. \editmark{The system generates captions based on the structural and spatial relations of the graphical objects and their properties along with explanations describing the perceptual complexity of the data attributes in the graphics.} SumTime~\cite{yu:2003} used pattern recognition techniques to
generate textual summaries of time-series data. The iGRAPH-Lite system~\cite{ferres:2007} 
made information in a graphic accessible to blind users by using
templates to provide textual summaries of what the graphic looks
like. The summaries however, do not focus on the higher-level takeaway
conveyed by the graphic. Chen et
al.~\cite{chen2019figure} produced natural language descriptions for
figures by identifying relations among labels present in the
figures.

Other work has explored natural language generation techniques
for assembling multiple caption units together to form
captions~\cite{qian2020formative}. Deep learning techniques based on
neural networks automate caption generation tasks for news
images~\cite{chen2019neural}.  Elzer et
al.~\cite{elzer2011automated} identified communicative signals that
represent the intent of messages portrayed in basic bar charts by
applying a Bayesian network methodology for reasoning about these
signals and generating captions. \editmark{Liu et
  al.}~\cite{shixia:2009, wei:2010,shixia:2012} explored the
integration of text analytics algorithms with interactive
visualization tools to help users understand and interpret the
summarization results. Contexifier~\cite{hullman2013contextifier}
automatically annotated visualizations of stock behavior with news
article headlines taking into account visual salience contextual
relevance, and key events from the articles. Kim et
al.~\cite{kim2020answering} introduced an automatic chart question
answering pipeline that generates visual explanations that refer to
visual features of charts using a template-based natural language
generation approach. \editmark{Voder~\cite{voder} generated data facts
  for visualizations with embellishments to help users interpret
  visualizations and communicate their findings. While an evaluation
  of that system suggested that interactive data facts aided users in
  interpreting visualizations, the paper did not specifically explore
  the interplay between data facts and the visualizations and their
  effects on the readers' takeaways.}

\editmark{These systems focus on helping authors with auto-generated text that can be associated with graphics; however, the work does not evaluate what information readers gather from the generated captions with their corresponding graphics.  Our paper specifically explores how
similarities and differences between what is visually emphasized in a line chart and textually
emphasized in its caption, can affect what readers take
away from the information when presented together. Future directions from our work could extend the functionality of chart authoring tools by providing automatic suggestions for captions as well as for chart presentation to help the reader take away information that is consistently emphasized by both the chart and caption.}

\section{Study}

\singlecol{
\begin{wrapfigure}{R}{0.5\textwidth}
  \includegraphics[width=\linewidth]{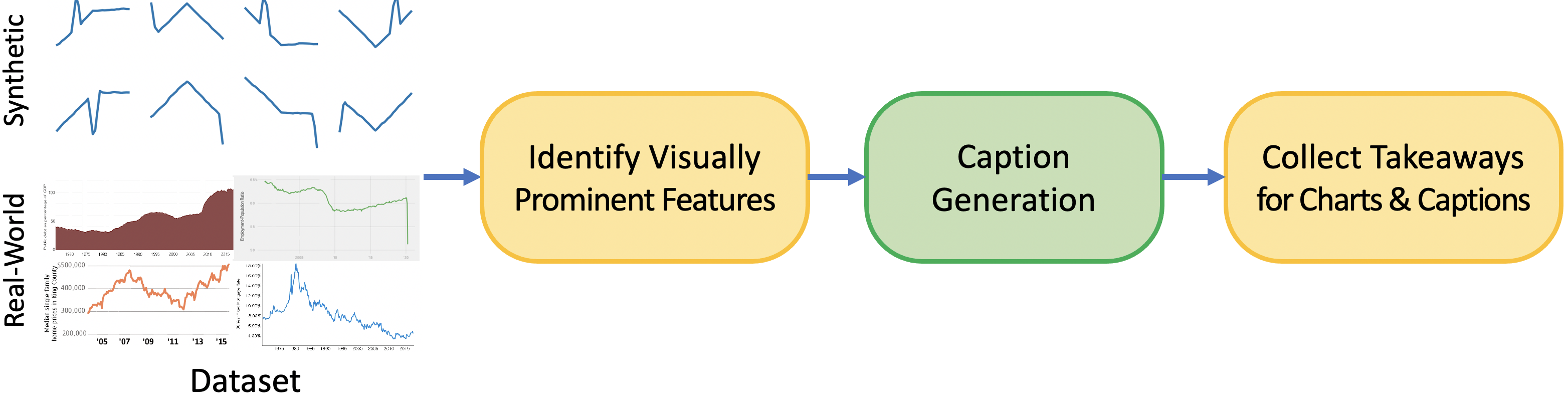}
  \caption{Our study pipeline. The inputs to the study are $27$ synthetic and $16$ real-world charts. Yellow boxes represent steps where we employed crowdsourcing. \editmark{The green box indicates that the step did not involve crowdsourcing.}}
  \label{fig:pipeline}
  \Description[Our study pipeline.]{Our study pipeline. The inputs to the study are 27 synthetic and 16 real-world charts. Yellow boxes represent steps where we employed crowdsourcing. The green box indicates that the step did not involve crowdsourcing.}
\end{wrapfigure}
}

\doublecol{
\begin{figure}
  \includegraphics[width=\linewidth]{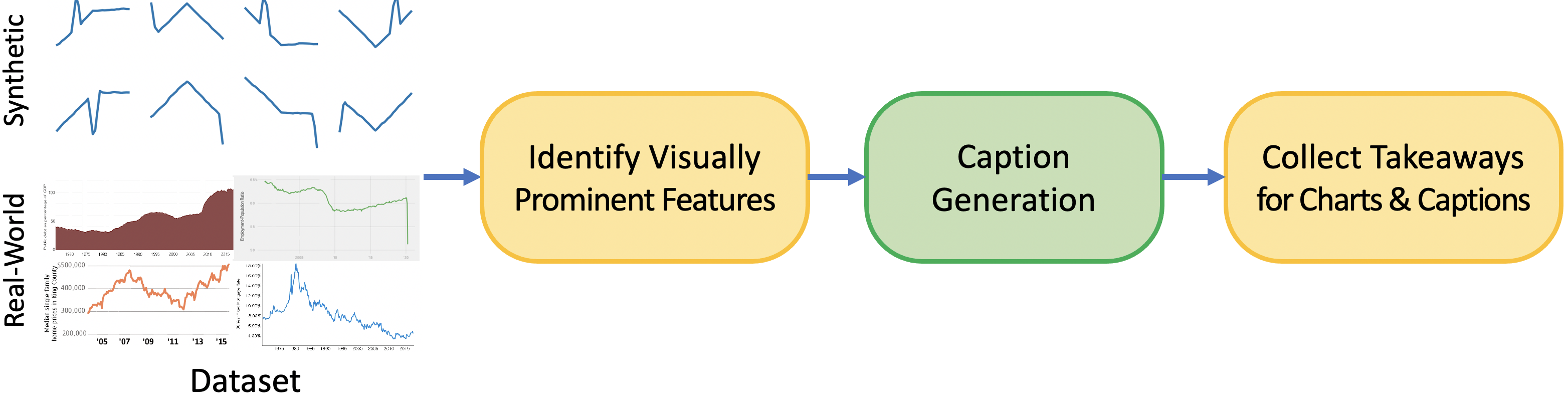}
  \caption{Our study pipeline. The inputs to the study are $27$ synthetic and $16$ real-world charts. Yellow boxes represent steps where we employed crowdsourcing.  \editmark{The green box indicates that the step did not involve crowdsourcing.}}
  \label{fig:pipeline}
  \Description[Our study pipeline.]{Our study pipeline. The inputs to the study are 27 synthetic and 16 real-world charts. Yellow boxes represent steps where we employed crowdsourcing. The green box indicates that the step did not involve crowdsourcing.}
  \vspace{-5mm}
\end{figure}
}

\editmark{We conducted a crowdsourced study to understand how captions describing features of varying prominence levels and the effect of including or not including external information for context, interacts with the chart in forming the readers' takeaways}. Through an initial data collection activity, we asked participants to identify features in the line charts that they thought were visually prominent. We generated captions corresponding to those marked features of various levels of prominence. We then ran a study asking a new set of participants to type their takeaways after viewing a chart and caption pair. Figure~\ref{fig:pipeline} shows the study pipeline.

\subsection{Datasets}

\singlecol{
\begin{figure}[ht]
  \includegraphics[width=\textwidth]{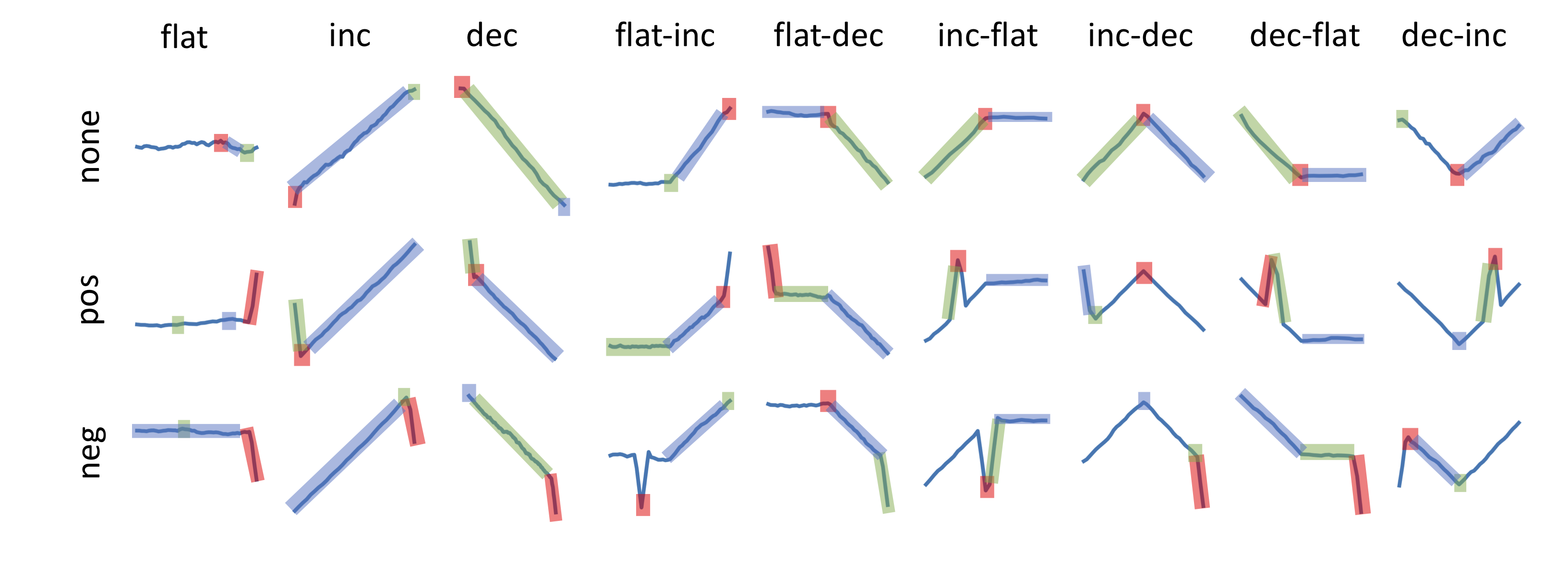}
  \caption{The $27$ data shapes generated for the study and their top three prominent features. Columns represent the nine possible global shapes and rows represent the three possible local outlier types. Here, `flat', `inc', and `dec' denote flat, increasing, and decreasing trends respectively. `none', `neg', and `pos' denote none, negative, and positive outlier types respectively. Red, green, and blue regions indicate the top three prominent features in order.}
  \label{fig:shape-matrix}
  \Description[The 27 data shapes with top three prominent features.]{The 27 data shapes generated for the study and their top three prominent features. Columns represent the nine possible global shapes and rows represent the three possible local outlier types. Here, `flat', `inc', and `dec' denote flat, increasing, and decreasing trends respectively. `none', `neg', and `pos' denote none, negative, and positive outlier types respectively. Red, green, and blue regions indicate the top three prominent features in order.}
\end{figure}  
}
\doublecol{
\begin{figure*}[ht]
  \includegraphics[width=\textwidth]{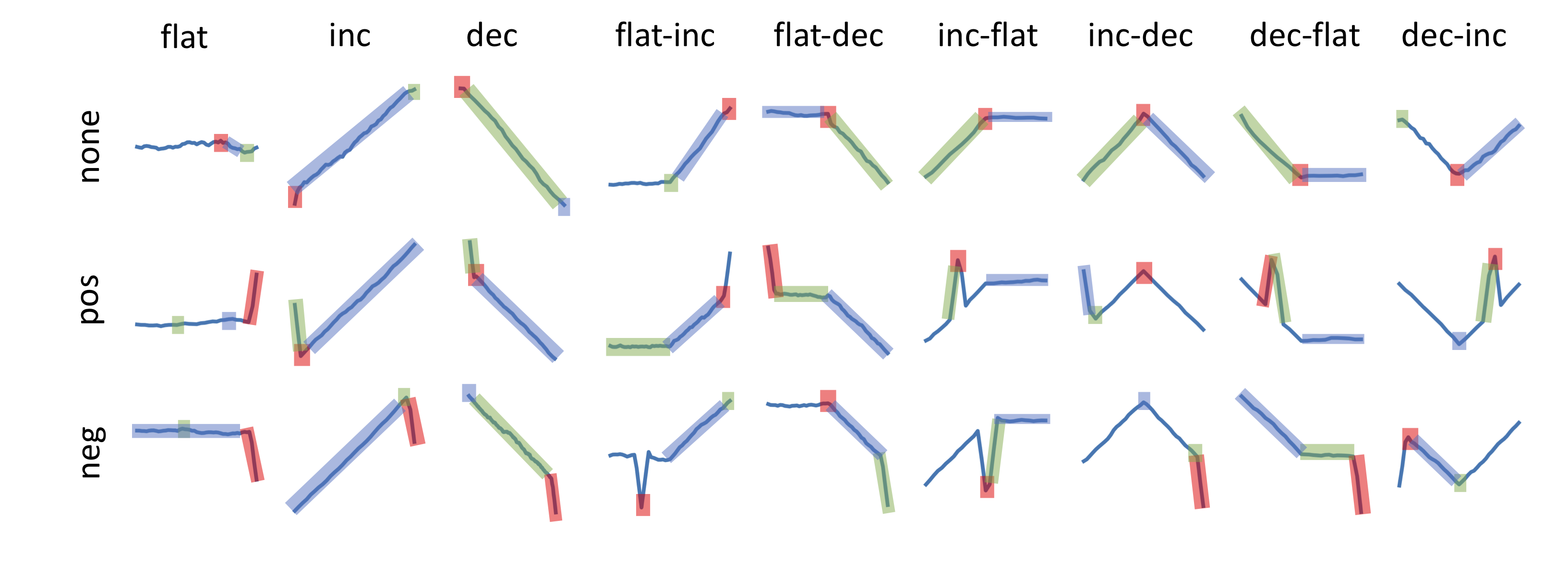}
  \vspace{-12mm}
  \caption{The $27$ data shapes generated for the study and their top three prominent features. Columns represent the nine possible global shapes and rows represent the three possible local outlier types. Here, `flat', `inc', and `dec' denote flat, increasing, and decreasing trends respectively. `none', `neg', and `pos' denote none, negative, and positive outlier types respectively. Red, green, and blue regions indicate the top three prominent features in order.}
  \label{fig:shape-matrix}
  \Description[The 27 data shapes with top three prominent features.]{The 27 data shapes generated for the study and their top three prominent features. Columns represent the nine possible global shapes and rows represent the three possible local outlier types. Here, `flat', `inc', and `dec' denote flat, increasing, and decreasing trends respectively. `none', `neg', and `pos' denote none, negative, and positive outlier types respectively. Red, green, and blue regions indicate the top three prominent features in order.}
  \vspace{-2mm}
\end{figure*}  
}

\singlecol{
\begin{figure}[ht]
  \includegraphics[width=\linewidth]{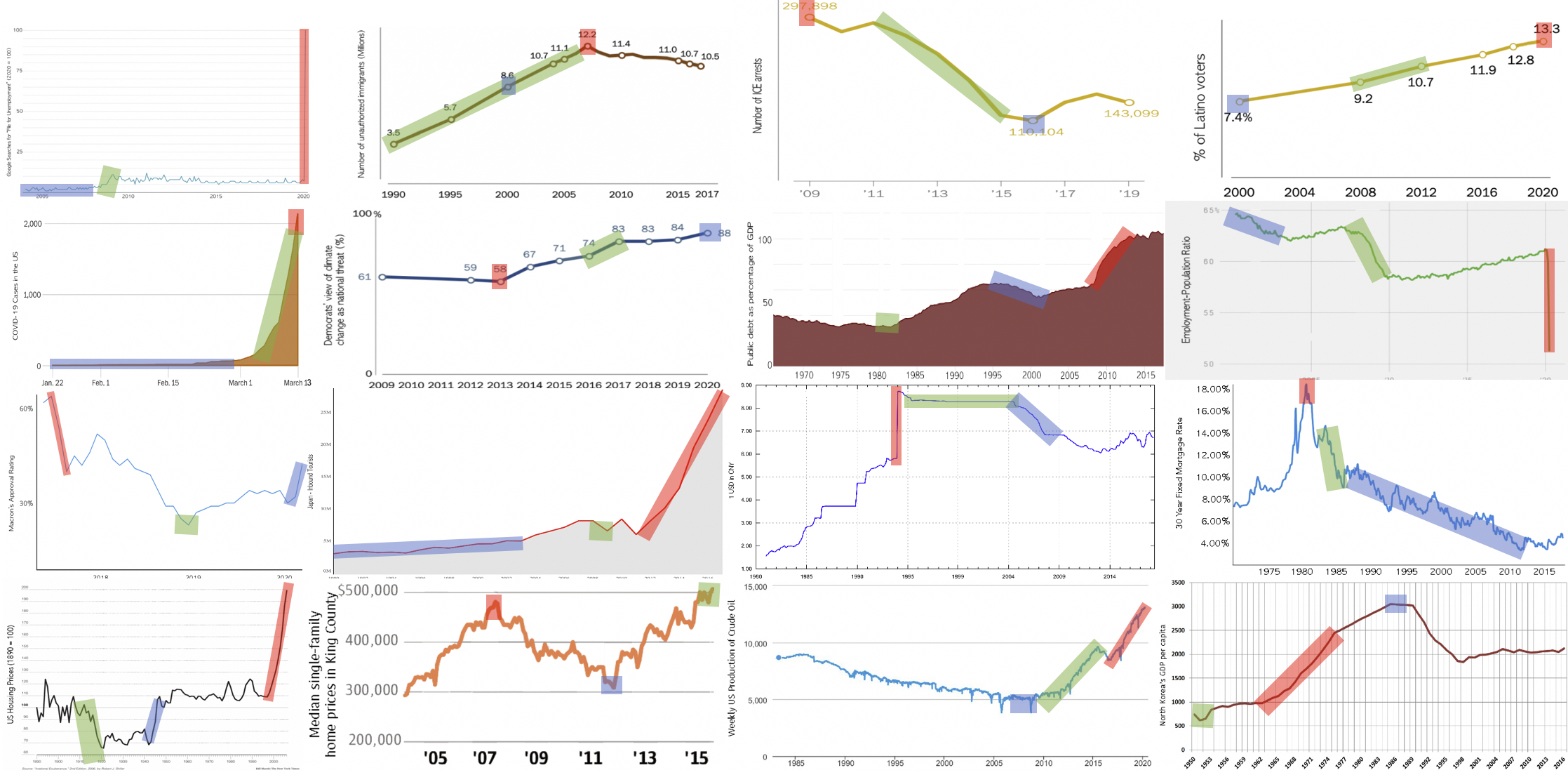}
  \caption{The $16$ real-world charts. Red, green, and blue regions indicate the top three prominent features in order.}
  \label{fig:charts-wild}
  \Description[The 16 real-world charts with top three prominent features.]{The 16 real-world charts. Red, green, and blue regions indicate the top three prominent features in order.}
  \end{figure}  
}
\doublecol{
\begin{figure*}[ht]
  \includegraphics[width=\linewidth]{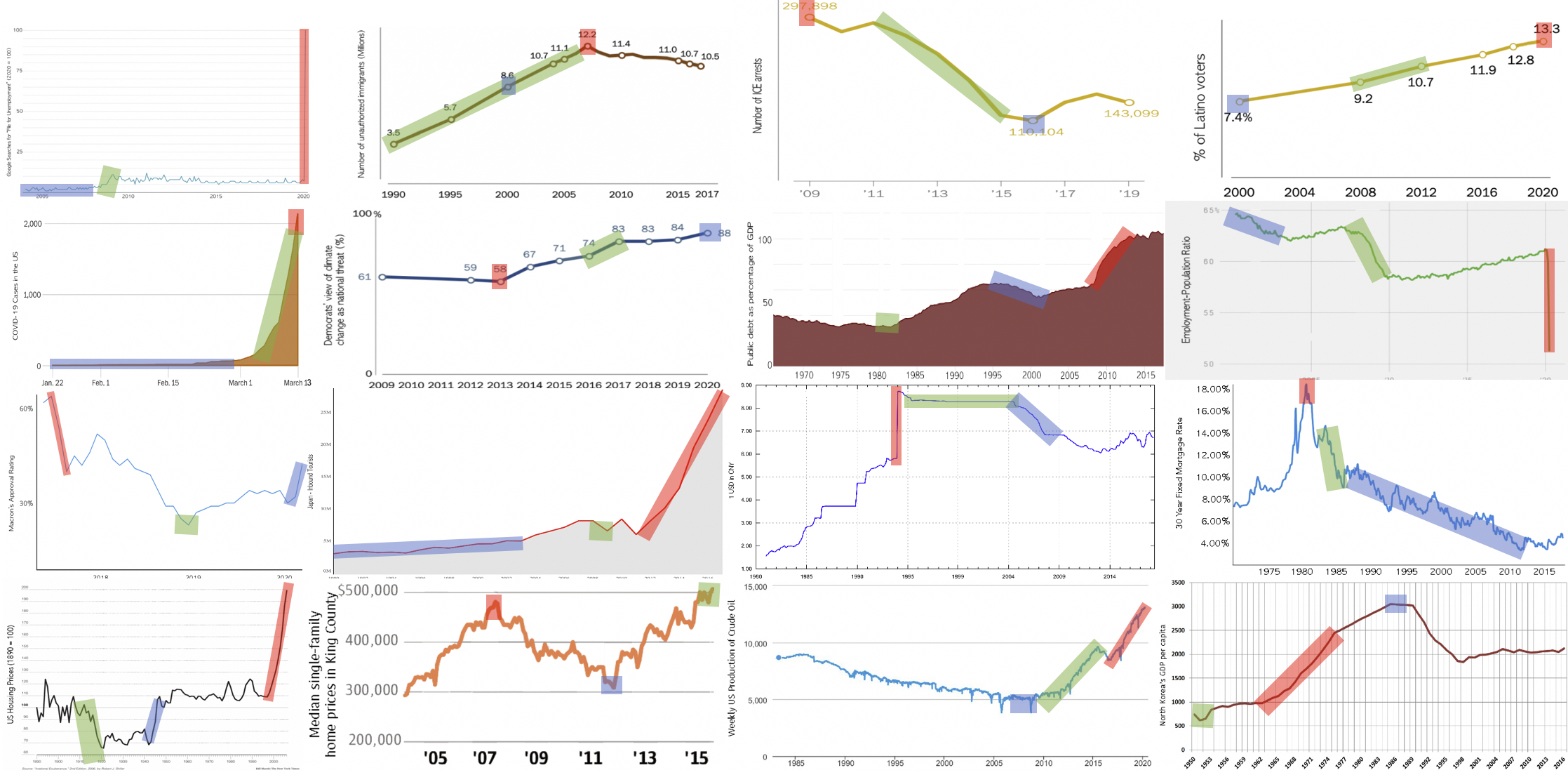}
  \vspace{-7mm}
  \caption{The $16$ real-world charts. Red, green, and blue regions indicate the top three prominent features in order.}
  \label{fig:charts-wild}
  \Description[The 16 real-world charts with top three prominent features.]{The 16 real-world charts. Red, green, and blue regions indicate the top three prominent features in order.}
  \vspace{-3mm}
  \end{figure*} 
}

We ran the study on two different datasets - (1) synthetically generated line charts that we designed to ensure good coverage of a variety of visual features that occur in line charts
and (2) line charts gathered from real-world sources to
serve as a more ecologically valid setting for our study.

\vspace{0.05in}
\noindent {\bf \em Synthetic Charts.}
We generated a set of synthetic line charts with
common visual features (i.e., trends, extrema, and
inflection points) while maintaining realistic global shapes.
To keep the overall design space tractable, we limited 
global shapes to include at most two trends (i.e., up, down, and
flat) and added at most one perturbation to induce features (e.g. inflection points) in either the positive or negative direction, 
resulting in a total of $27$ data shapes
(Figure~\ref{fig:shape-matrix}).  To provide context to the charts, we labeled the x-axis with time unit values implying that the chart represents a time series. Specifically, we
selected the start and end of the x-axis from the set of years \{1900,
1910, 1920,..., 2020\}. To label the y-axis, we chose a domain for the y-axis and its value range
from the MassVis dataset~\cite{borkin2013makes}.

\vspace{0.05in}
\noindent {\bf \em Real-world Charts.}  To build a more ecologically
representative dataset of line charts with various shapes, styles, and
domains, we collected $16$ charts (Figure~\ref{fig:charts-wild}) from sources such as The
Washington Post~\cite{washingtonpost}, Pew Research~\cite{pewresearch}, Wikipedia~\cite{wikipedia}, and Tableau Public~\cite{tableaupublic}. Because our study focuses on prominence arising from intrinsic features in line charts, we removed all graphical elements that could potentially affect the
prominence of the features in the charts (e.g., text annotations, highlighting, and background shading).
In addition, we removed all text except for the axis labels (e.g. chart titles) so that the
captions serve as the primary source of text provided with the chart. We added axis labels to those charts without labels to ensure readability.

\subsection{Identify Visually Prominent Features}
\singlecol{
\begin{figure}
  \includegraphics[width=\textwidth]{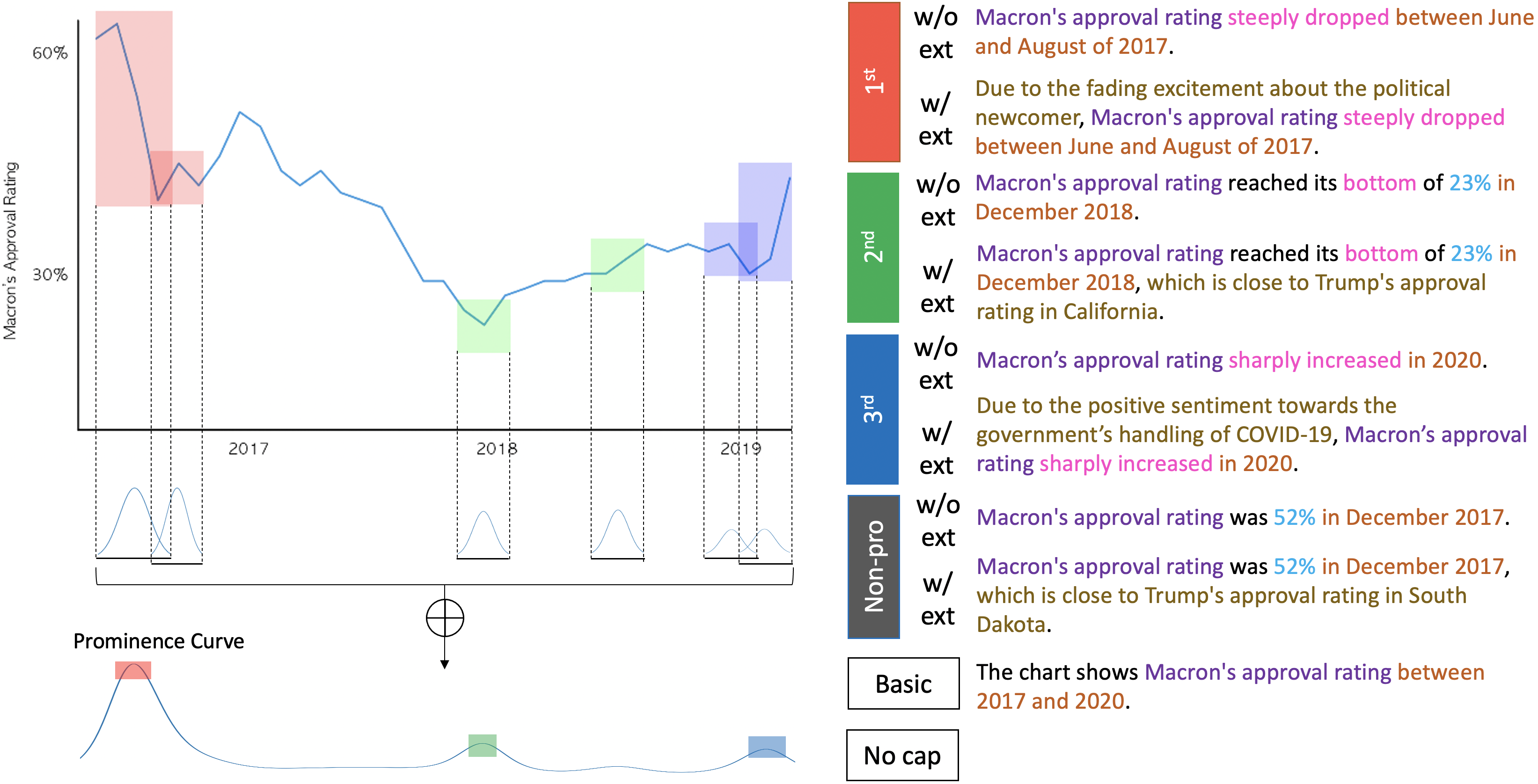}
  \caption{The line on the bottom \editmark{left} shows the prominence curve for the line chart above. From this curve, we obtain the most prominent (red), the second most prominent (green), and the third most prominent (blue) features in the chart. The $10$ caption variants (one of them being a no-caption variant) generated based on these prominent features, are shown on the right. The text colors indicate the types of fill-in values based on the caption templates; \purple{purple} for dimensions, \fuchsia{fuchsia} for the feature description, \blue{blue} for data values, and \brown{brown} for the time period.}
  \label{fig:voting-caption}
  \Description[Generation of 10 caption variants based on the prominence curve.]{The line on the bottom left shows the prominence curve for the line chart above. From this curve, we obtain the most prominent (red), the second most prominent (green), and the third most prominent (blue) features in the chart. The 10 caption variants (one of them being a no-caption variant) generated based on these prominent features, are shown on the right. The text colors indicate the types of fill-in values based on the caption templates; purple for dimensions, fuchsia for the feature description, blue for data values, and brown for the time period.}
\end{figure}
}
\doublecol{
\begin{figure*}[t]
  \includegraphics[width=\textwidth]{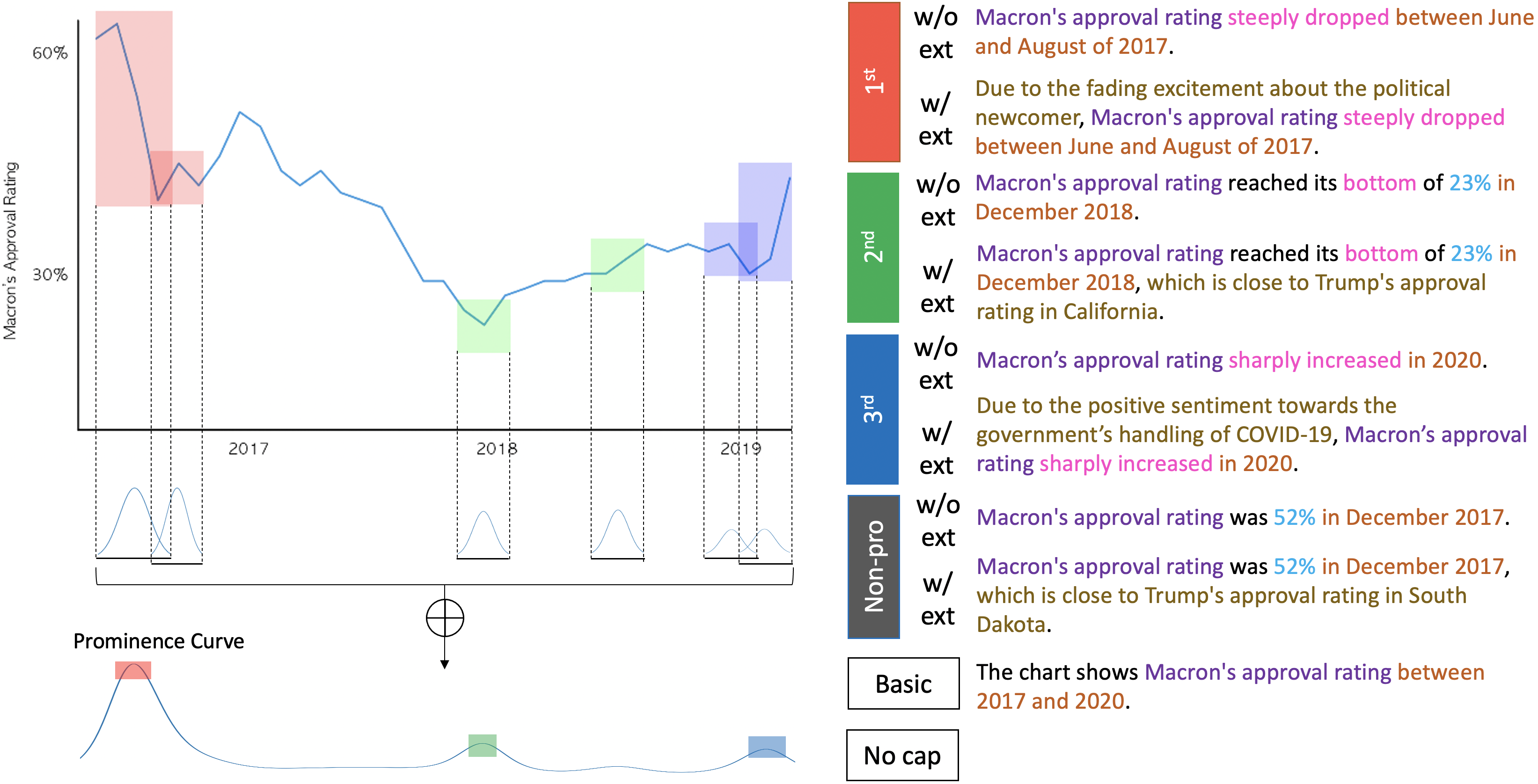}
  \vspace{-5mm}
  \caption{The line on the bottom \editmark{left} shows the prominence curve for the line chart above. From this curve, we obtain the most prominent (red), the second most prominent (green), and the third most prominent (blue) features in the chart. The $10$ caption variants (one of them being a no-caption variant) generated based on these prominent features, are shown on the right. The text colors indicate the types of fill-in values based on the caption templates; \purple{purple} for dimensions, \fuchsia{fuchsia} for the feature description, \blue{blue} for data values, and \brown{brown} for the time period.}
  \label{fig:voting-caption}
  \Description[Generation of 10 caption variants based on the prominence curve.]{The line on the bottom left shows the prominence curve for the line chart above. From this curve, we obtain the most prominent (red), the second most prominent (green), and the third most prominent (blue) features in the chart. The 10 caption variants (one of them being a no-caption variant) generated based on these prominent features, are shown on the right. The text colors indicate the types of fill-in values based on the caption templates; purple for dimensions, fuchsia for the feature description, blue for data values, and brown for the time period.}
  \vspace{-3mm}
\end{figure*}
}

To identify the most visually prominent features in our dataset, we recruited at least five workers from Amazon Mechanical Turk~\cite{turk} for each line chart and asked them to draw rectangular bounding boxes around the top three most prominent features in the chart. We also asked them to
briefly describe each marked feature in their own words so that we
could differentiate between trend and slope features versus peak, inflection, and other point features.

\editmark{In each trial of the data collection, we presented one of the $43$ line charts.} Because we were seeking subjective responses, each participant completed only one trial to avoid
biases that might arise from repeated exposure to the
task. Participation was limited to English speakers in the U.S. with
at least a 98\% acceptance rate and $5000$ approved tasks. We payed a
rate equivalent to \$2 / 10 mins.

\editmark{We asked a total of $219$ participants (average of $5.09$ per chart) to label the top three features for a total of $657$ prominence boxes.}
\cut{We collected $408$ prominence boxes from $136$ participants for the
synthetic charts and $249$ prominence boxes from $83$ participants for
the real-world charts, with a total of $219$ participants (average of
$5.09$ per chart) labeling the top three features.}
We then aggregated all of the feature bounding boxes provided by first
projecting each box onto the x-axis, to form a 1D interval
(Figure~\ref{fig:voting-caption} \editmark{upper left}). We weighted each interval inversely proportional to
the ranking provided by the participant. Specifically, the top ranked feature
bounding box for each participant was assigned a weight of $3$, while the 3rd
ranked feature was assigned a weight of $1$. \editmark{We noticed that bounding boxes corresponding to the same features were pretty consistent in the central regions although the exact boundary drawn by the participants varied. In order to boost the signal in the central regions while suppressing the noise in the boundary regions, we multiplied \editmark{the weight assigned to} each
interval by a Gaussian factor centered at the interval
and with standard deviation set to half the width of the interval.} Summing all of the Gaussian weighted intervals, we obtained a {\em prominence curve} \editmark{(Figure~\ref{fig:voting-caption} bottom left)}. \editmark{However, a region defined by a local maximum of the curve may not have an obvious one-to-one mapping with a feature in the chart because it roughly indicates a high prominence region instead of pinpointing a specific visual feature. We considered all the bounding boxes containing the region along with the participants' text descriptions of the features to associate the local maximum to a certain feature.} We iterated this process \editmark{for the region around the top three local maximum to} identify three prominent features. \editmark{Results of the algorithm for the charts in our dataset are shown in Figures~\ref{fig:shape-matrix} and \ref{fig:charts-wild}.} 

\subsection{Caption Generation}
\singlecol{
\begin{wraptable}{R}{0.55\textwidth}
\begin{center}
\begin{tabular}{ |l|l| }
 \hline
 Feature & Template \\
 \hline
 {\footnotesize Extremum} & {\scriptsize \purple{\texttt{[dimension]}} \textit{reached its} \fuchsia{\texttt{[extrema-word]}} \textit{of} \blue{\texttt{[value]}} \textit{in} \brown{\texttt{[time-period]}}.} \\
 {\footnotesize Trend} & {\footnotesize \purple{\texttt{[dimension]}} \fuchsia{\texttt{[slope-word]}}  \textit{in | between} \brown{\texttt{[time-period]}}.} \\
 {\footnotesize Inflection} & {\footnotesize \purple{\texttt{[dimension]}} \textit{started} \fuchsia{\texttt{[slope-word]}} \textit{in} \brown{\texttt{[time-period]}}.} \\
 {\footnotesize Point} & 
{\footnotesize \purple{\texttt{[dimension]}} \textit{was} \blue{\texttt{[value]}} \textit{in} \brown{\texttt{[time-period]}}.} \\
 \hline
\end{tabular}
\end{center}
\caption{\editmark{Examples of templates we employed for generating captions about specific features. The text colors indicate the types of fill-in values based on
the caption templates; \purple{purple} for dimensions, \fuchsia{fuchsia}
for feature descriptions, \blue{blue} for data values, and
\brown{brown} for time periods. Examples of filled in captions are in Figure~\ref{fig:voting-caption} (right).}}
  \label{tab:caption-templates}
\end{wraptable}
}

\editmark{To carefully control the language used in the captions and
  keep the number of conditions manageable, we generated captions
  using templates that only vary the feature mentioned and whether external information is introduced.
  Using the templates, we produced the following caption variants: (1)
  two captions (one with and one without external information) for each of the top three visually prominent features identified earlier, (2)
  two captions (one with and one without external information) describing a
  minimally prominent feature that is neither an extremum nor an
  inflection point, and (3) a basic caption that simply describes the
  domain represented in the chart without describing a particular
  feature.}

We generated $10$ caption variants (including the no caption variant in which we presented a chart without caption) for each of the $43$
charts, providing a total of $430$ chart-caption pairs.
We \editmark{manually} generated all the captions rather than using the original captions for the real-world
charts to control for word use and grammatical structure. \editmark{For real-world charts, we searched for information from the document that they originally appeared in, to extract information not present in the charts. In particular, we looked for information about potential reasons for trends or change (e.g. the external information included in the caption about the most prominent feature in Figure~\ref{fig:voting-caption}) or comparisons with a similar entity (e.g. comparison between Macron's approval rating with Trump's approval rating in the second most prominent feature in Figure~\ref{fig:voting-caption}).} 
For \editmark{synthetically generated charts and real-world charts
  that were not accompanied with additional information about their
  features,} we referenced Wikipedia~\cite{wikipedia} articles to
create a plausible context. 

\editmark{We employed simple language templates for caption generation
  to minimize the effects of linguistic variation
  (Table~\ref{tab:caption-templates}). The captions generated with the
  templates were allowed to vary in the features they described in the
  charts. To make the descriptions of the features appear natural, we
  included words the participants used to describe the features during the
  prominent feature collection phase. Because the participants usually
  described each of the features using a noun occasionally with an
  adjective modifier (e.g. ``sharp increase''), we manually
  lemmatized the words and modified the forms to correctly fit into
  our template (e.g. ``sharply increased'' in the caption
  about the third most prominent figure in
  Figure~\ref{fig:voting-caption}).}

\doublecol{
\begin{table}
\begin{center}
\begin{tabular}{ |l|l| }
 \hline
 \footnotesize{Feature} & \footnotesize{Template} \\
 \hline
 {\footnotesize Extremum} & {\scriptsize \purple{\texttt{[dimension]}} \textit{reached its} \fuchsia{\texttt{[extrema-word]}} \textit{of} \blue{\texttt{[value]}} \textit{in} \brown{\texttt{[time-period]}}.} \\
 {\footnotesize Trend} & {\footnotesize \purple{\texttt{[dimension]}} \fuchsia{\texttt{[slope-word]}}  \textit{in | between} \brown{\texttt{[time-period]}}.} \\
 {\footnotesize Inflection} & {\footnotesize \purple{\texttt{[dimension]}} \textit{started} \fuchsia{\texttt{[slope-word]}} \textit{in} \brown{\texttt{[time-period]}}.} \\
 {\footnotesize Point} & 
{\footnotesize \purple{\texttt{[dimension]}} \textit{was} \blue{\texttt{[value]}} \textit{in} \brown{\texttt{[time-period]}}.} \\
 \hline
\end{tabular}
\end{center}
\caption{\editmark{Examples of templates we employed for generating captions about specific features. The text colors indicate the types of fill-in values based on
the caption templates; \purple{purple} for dimensions, \fuchsia{fuchsia}
for feature descriptions, \blue{blue} for data values, and
\brown{brown} for time periods. Examples of filled in captions are in Figure~\ref{fig:voting-caption} (right).}}
  \label{tab:caption-templates}
  \vspace{-9mm}
\end{table}
}

\subsection{Collect Takeaways for Charts \& Captions}

\doublecol{
\begin{figure}
  \includegraphics[width=\linewidth]{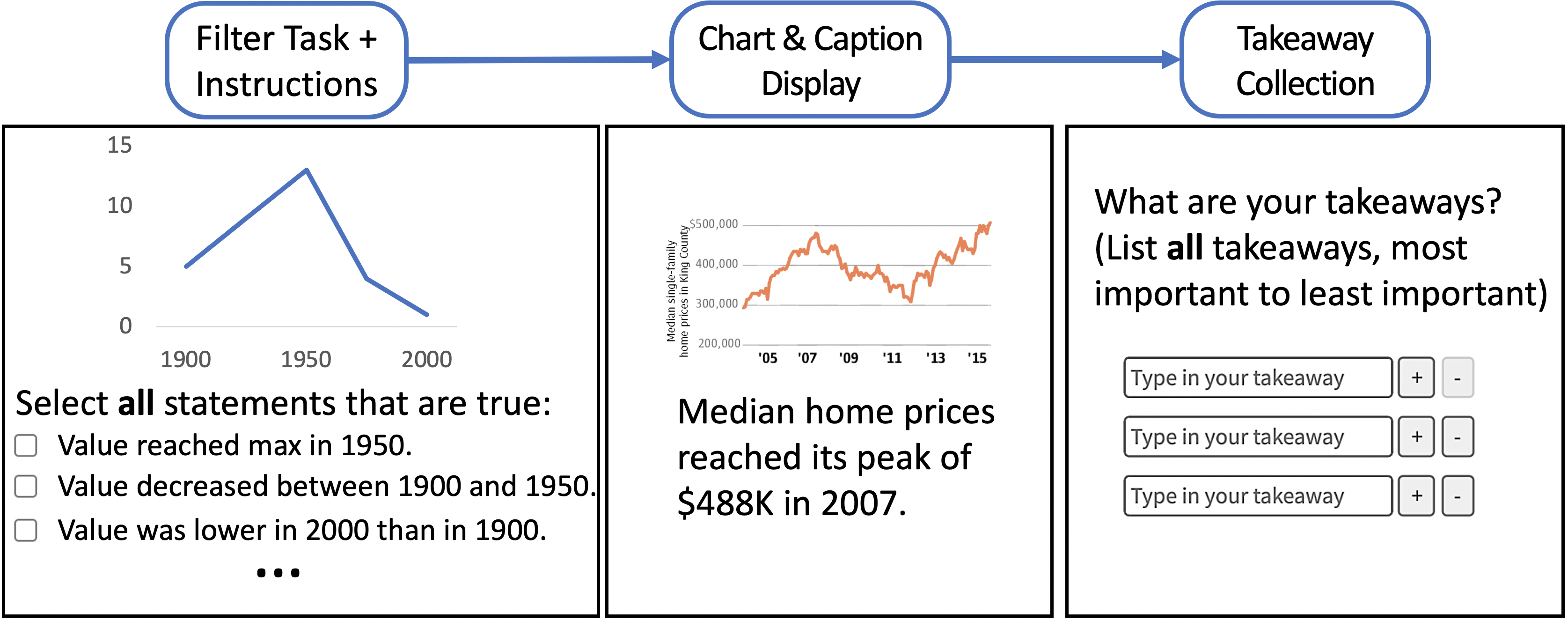}
  \vspace{-3mm}
  \caption{\editmark{The procedure for collecting takeaways for chart-caption pairs. The images show simplified versions of the screen that the participants saw during each step.}}
  \label{fig:study-screens}
  \Description[The procedure for collecting takeaways for chart-caption pairs.]{The procedure for collecting takeaways for chart-caption pairs. The images show simplified versions of the screen that the participants saw during each step.}
  \vspace{-3mm}
\end{figure}
}

\subsubsection{Design}
We ran a between-subjects design study for collecting takeaways for charts and their captions.
For each of the $43$ charts, we presented one of the \editmark{ten variants (including the no caption variant)} (\editmark{examples in} Figure~\ref{fig:voting-caption}):\\\\[3pt] 
{\small
(1) [1st w/o ext] Caption for most prominent feature, no external info.\\[1pt]
(2) [1st w/ ext] Caption for most prominent feature, has external info.\\[1pt]
(3) [2nd w/o ext] Caption for 2nd most prominent feature, no external info.\\[1pt]
(4) [2nd w/ ext] Caption for 2nd most prominent feature, has external info.\\[1pt]
(5) [3rd w/o ext] Caption for 3rd most prominent feature, no external info.\\[1pt]
(6) [3rd w/ ext] Caption for 3rd most prominent feature, has external info.\\[1pt]
(7) [non-pro w/o ext] Caption for non-prominent feature, no external info.\\[1pt]
(8) [non-pro w/ ext] Caption for non-prominent feature, has external info.\\[1pt]
(9) [basic] Caption about domain represented in the chart and $x$-range\\[1pt]
(10) [no cap] No caption
}

\subsubsection{Procedure}
\singlecol{
\begin{wrapfigure}{R}{0.5\textwidth}
  \includegraphics[width=\linewidth]{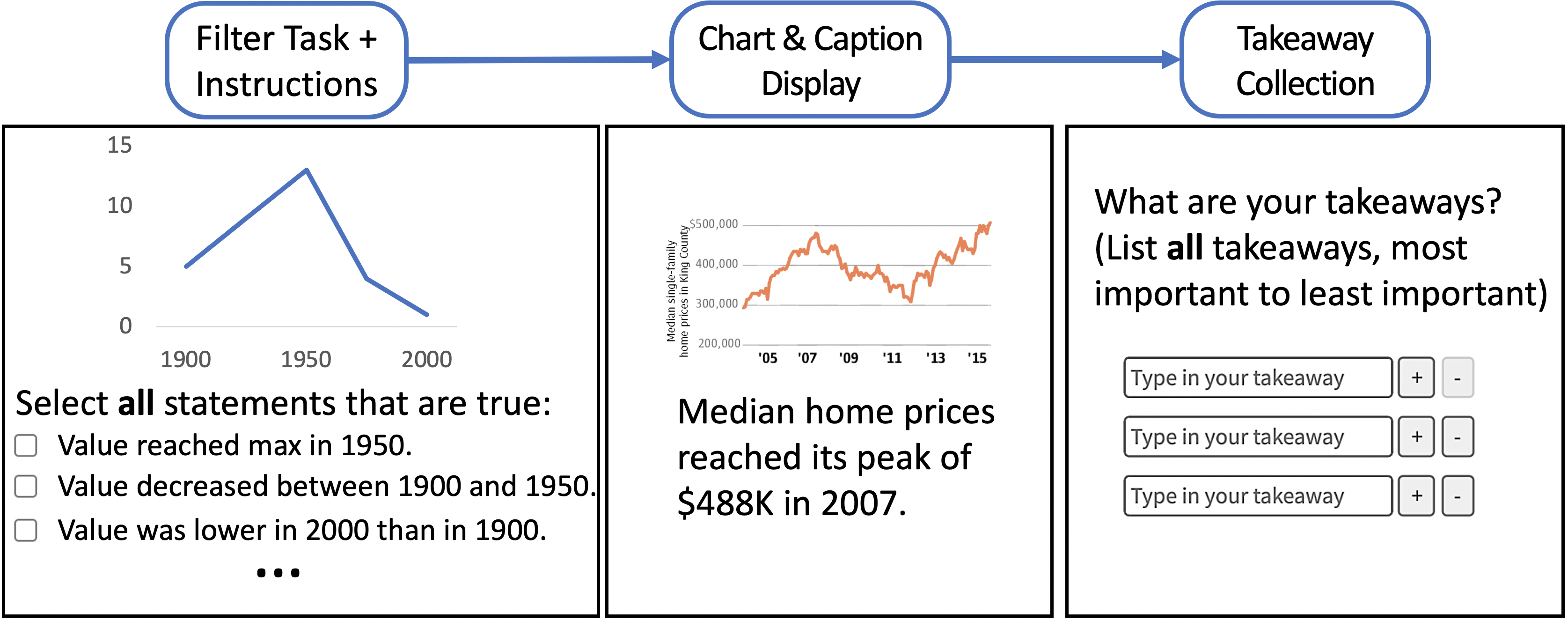}
  \caption{\editmark{The procedure for collecting takeaways for chart-caption pairs. The images show simplified versions of the screen that the participants saw during each step.}}
  \label{fig:study-screens}
  \Description[The procedure for collecting takeaways for chart-caption pairs.]{The procedure for collecting takeaways for chart-caption pairs. The images show simplified versions of the screen that the participants saw during each step.}
\end{wrapfigure}
}

The study began with a screening test to ensure that the participant
had a basic understanding of line charts and could \editmark{read values and encodings, extract extrema and trends, and compare values (Figure~\ref{fig:study-screens} first step)}. Only participants who passed this test were allowed to
continue with the study. After they read the instructions, the participants were presented with a chart
and a caption \editmark{underneath the chart, similar to most charts in
  the real world} (unless it is the no-caption variant) (Figure~\ref{fig:study-screens} second step). 
  \editmark{We
    did not impose a time constraint on the amount of time spent looking at the chart and the caption to allow participants sufficient time to read and digest the information at their own pace, like document reading in the real world.} \editmark{On the next screen for collecting takeaways, the chart and the caption were removed to constrain readers to provide the takeaways based on memory instead
  of simply re-reading from the chart and the caption.}
  The participants were asked to list as many text
takeaways as they could in the \editmark{order of
  importance (Figure~\ref{fig:study-screens} third step)}. 
 Finally, using
a 5-point Likert scale, we asked how much they relied on the chart and
caption individually when determining their takeaways.

\editmark{We asked each participant to provide takeaways for exactly one chart-caption pair to prevent potential biases from already having read a different caption about a chart.} \editmark{From \editmark{$2168$} participants (average of \editmark{$5.04$} per chart-caption pair), we collected a total of \editmark{$4953$} takeaways (average of \editmark{$2.28$} per participant).}

\subsubsection{Labeling Takeaways}
In order to analyze the takeaways, we manually labeled each takeaway
with the corresponding chart feature described.  \editmarkvar{
  Since participants often described multiple chart features in a
  single takeaway, we first split each takeaway into separate takeaways for each 
  visual feature mentioned.
   At the end of this process, we identified on average $1.31$ features
  per takeaway.  If the referenced feature was one of three most prominent
  features or the non-prominent feature we identified during caption
  generation, we labeled the takeaway with the corresponding
  feature, otherwise
   we labeled the takeaway as referring to an
  \textit{other} feature. If the takeaway did not refer to
  any specific feature in the chart, we labeled the takeaway as a
  \textit{non-feature}. Examples of \textit{non-feature} takeaways
  include an extrapolation such as ``The value will continue to rise
  after 2020'' or a judgment such as ``I should buy gold" when looking
  at a chart showing the price of gold over time. One of the authors labeled the features and discussed any confusing cases with the other authors to converge on the final label.}


\section{Results}

\singlecol{
\begin{figure}[t!]
	\centering
	\includegraphics[width=\textwidth]{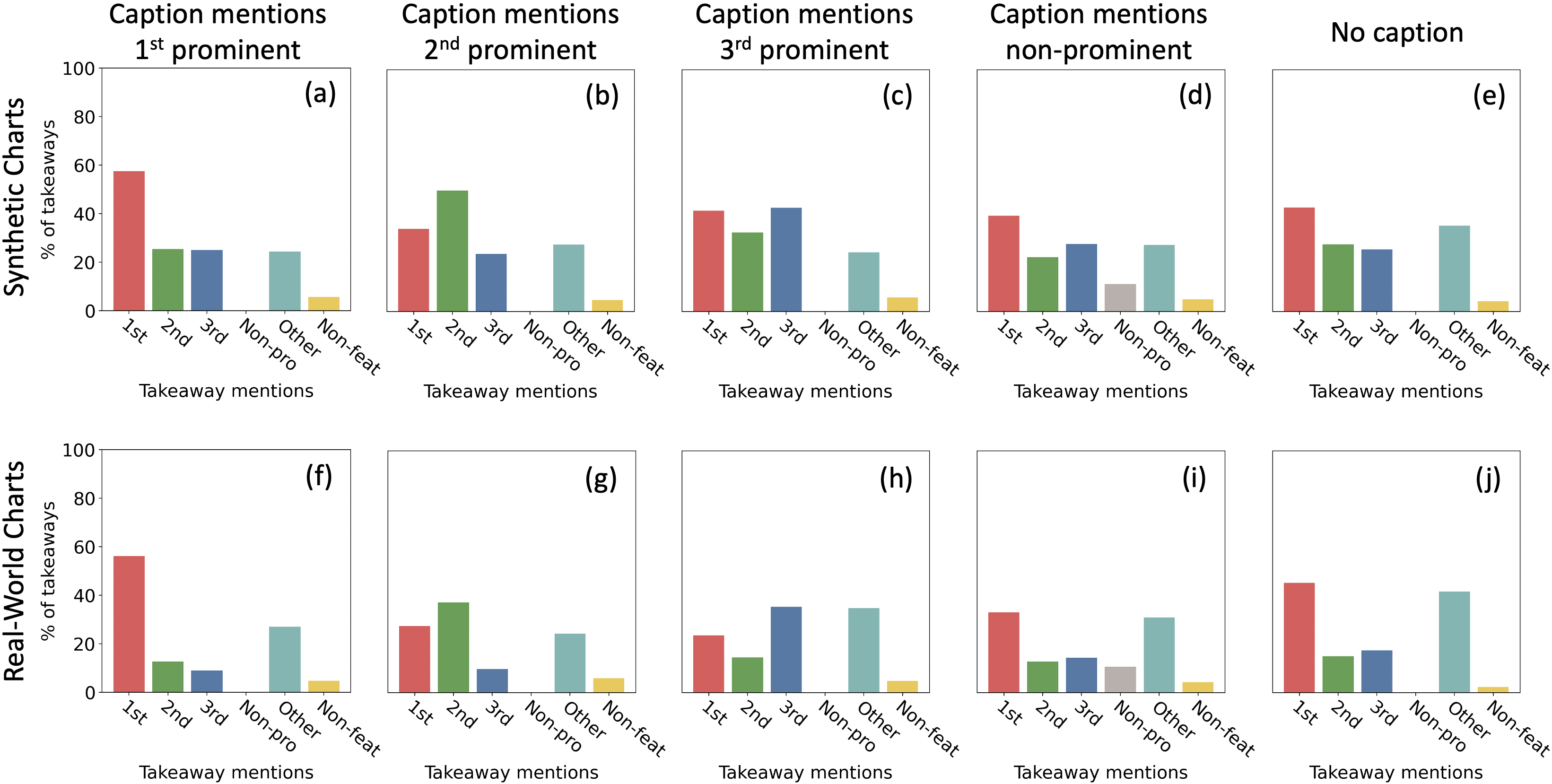}
        \caption{Study results. Each \editmark{column} shows bar charts for each prominence level
mentioned in the caption (i.e., the leftmost bar chart is for captions
mentioning the 1st ranked visual feature, the next bar chart is for
captions mentioning the 2nd ranked visual feature, while the rightmost
bar chart is for the no-caption condition).
Within a bar chart, each bar represents the percentage
of takeaways mentioning the
visual feature at that prominence level. For example, the
\editmark{leftmost} bar in each bar chart represents the percentage of
total takeaways that mention the top ranked takeaway. Each bar chart
also reports the percentage of {\em Other} features and {\em
  Non-features} that were mentioned in the takeaways. These
charts aggregate data for captions with and without external
information. \editmark{The percentages do not sum to $100\%$ as some takeaways mention multiple takeaways}.}
\vspace{-.25mm}
          \label{fig:h1-results}
          \Description[Study results.]{Study results. Each column shows bar charts for each prominence level mentioned in the caption (i.e., the leftmost bar chart is for captions
mentioning the 1st ranked visual feature, the next bar chart is for
captions mentioning the 2nd ranked visual feature, while the rightmost
bar chart is for the no-caption condition).
Within a bar chart, each bar represents the percentage
of takeaways mentioning the
visual feature at that prominence level. For example, the
leftmost bar in each bar chart represents the percentage of
total takeaways that mention the top ranked takeaway. Each bar chart
also reports the percentage of Other features and Non-features that were mentioned in the takeaways. These
charts aggregate data for captions with and without external
information. The percentages do not sum to $100\%$ as some takeaways mention multiple takeaways.}
\end{figure}
}
\doublecol{
\begin{figure*}[t!]
	\centering
	\includegraphics[width=\textwidth]{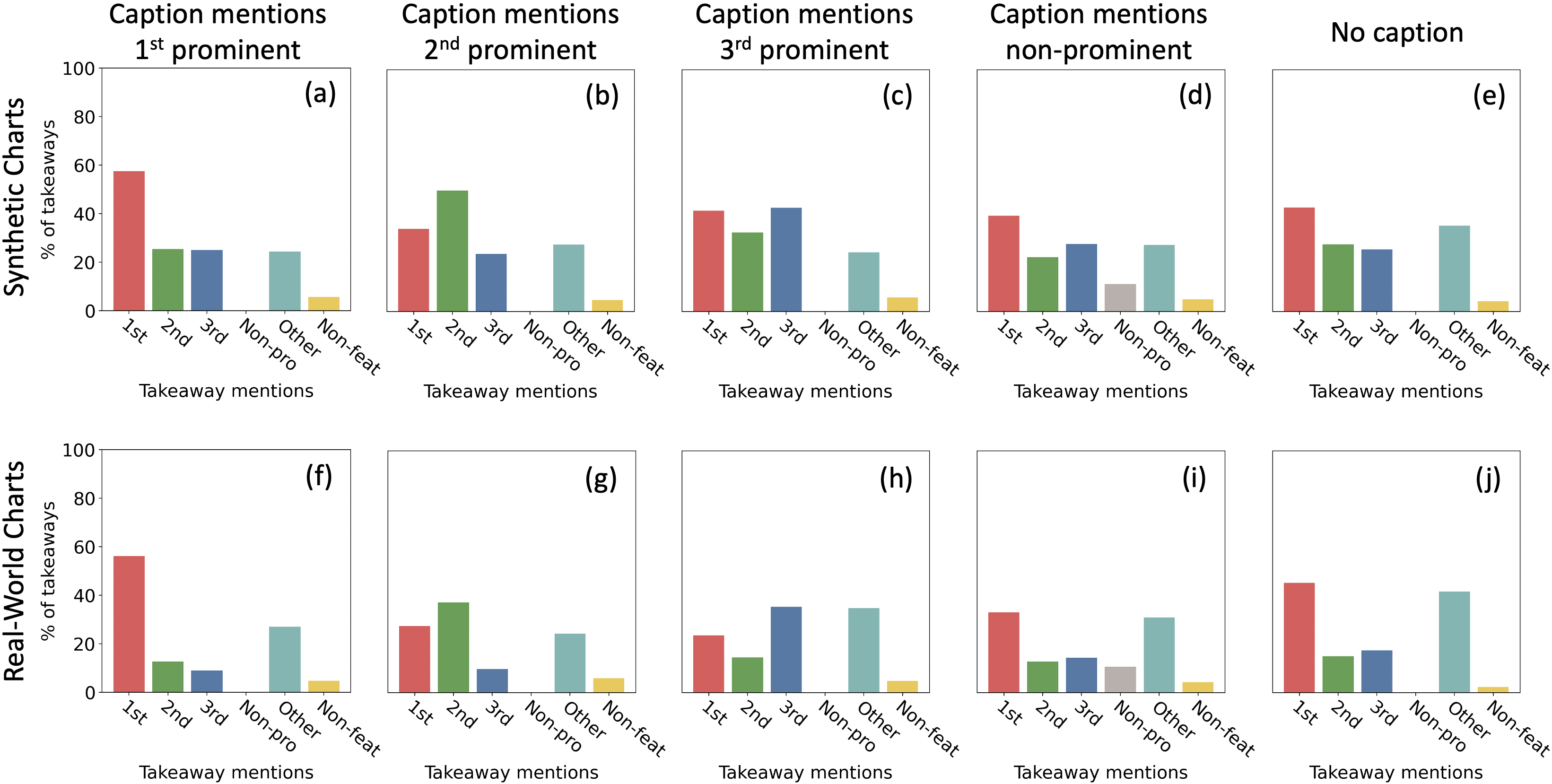}
	\vspace{-5mm}
        \caption{Study results. Each \editmark{column} shows bar charts for each prominence level
mentioned in the caption (i.e., the leftmost bar chart is for captions
mentioning the 1st ranked visual feature, the next bar chart is for
captions mentioning the 2nd ranked visual feature, while the rightmost
bar chart is for the no-caption condition).
Within a bar chart, each bar represents the percentage
of takeaways mentioning the
visual feature at that prominence level. For example, the
\editmark{leftmost} bar in each bar chart represents the percentage of
total takeaways that mention the top ranked takeaway. Each bar chart
also reports the percentage of {\em Other} features and {\em
  Non-features} that were mentioned in the takeaways. These
charts aggregate data for captions with and without external
information. \editmark{The percentages do not sum to $100\%$ as some takeaways mention multiple takeaways}.}
          \label{fig:h1-results}
          \Description[Study results.]{Study results. Each column shows bar charts for each prominence level mentioned in the caption (i.e., the leftmost bar chart is for captions
mentioning the 1st ranked visual feature, the next bar chart is for
captions mentioning the 2nd ranked visual feature, while the rightmost
bar chart is for the no-caption condition).
Within a bar chart, each bar represents the percentage
of takeaways mentioning the
visual feature at that prominence level. For example, the
leftmost bar in each bar chart represents the percentage of
total takeaways that mention the top ranked takeaway. Each bar chart
also reports the percentage of Other features and Non-features that were mentioned in the takeaways. These
charts aggregate data for captions with and without external
information. The percentages do not sum to $100\%$ as some takeaways mention multiple takeaways.}
\vspace{-3mm}
\end{figure*}
}

\cut{The primary goal of our study is to understand what readers take away when charts and captions
are presented together and how the emphasis on different prominent
features and presence of external information affects the takeaways. We analyze our results with respect to two hypotheses:
\begin{itemize}
\item {\bf H1:} When captions emphasize more visually prominent features of the chart, people are more likely to treat the features as the takeaway\editmark{; when a caption emphasizes a less visually prominent feature, people are less likely to treat that feature as the takeaway and more likely to treat a more visually prominent feature in the chart as the takeaway}.
\item {\bf H2:} When captions contain external information for context, \editmark{the external information} serves to further emphasize the feature presented in the caption and people are therefore more likely to treat that feature as the takeaway, compared to when the caption does not contain external information.
\end{itemize}
}
\editmark{The primary goal of our study is to understand what readers take away when charts and captions
are presented together and how the emphasis on different prominent
features and presence of external information affects the takeaways. We analyze our results with respect to two hypotheses: \\[3pt]
[{\bf H1}] When captions emphasize more visually prominent features of the chart, people are more likely to treat the features as the takeaway\editmark{; when a caption emphasizes a less visually prominent feature, people are less likely to treat that feature as the takeaway and more likely to treat a more visually prominent feature in the chart as the takeaway}. \\[1pt]
[{\bf H2}] When captions contain external information for context, \editmark{the external information} serves to further emphasize the feature presented in the caption and people are therefore more likely to treat that feature as the takeaway, compared to when the caption does not contain external information. \\[3pt]
}

\noindent {\bf \em Assessing H1.}
To evaluate {\bf H1}, we examine how varying the prominence of a visual
feature mentioned in a caption (independent variable), affects the
visual feature mentioned in the takeaways (dependent variable).  Figure~\ref{fig:h1-results} summarizes the
study results for the synthetic charts (top row) and the real-world charts
(bottom row).

In general, these results suggest that when a caption mentions visual
features of differing prominence levels, the takeaways also differ. 
Omnibus Pearson's chi-squared tests confirm a significant difference
between the bar charts for the 5 different caption conditions in both
the synthetic (\editmark{$\chi^{2} (20) = 202.211$}, \editmark{$p < 0.001$}) and real
world (\editmark{$\chi^{2} (20) = 207.573$}, \editmark{$p < 0.001$}) datasets. These results also suggest that when the caption mentions a specific feature, the takeaways also tend to mention that feature, when compared to
the baseline `no-caption' condition.

\singlecol{
\begin{table}
\begin{center}
\resizebox{0.7\linewidth}{!}{
\footnotesize{
\begin{tabular}{ |l|l|l|l|l||c|c| }
 \hline
 & \multicolumn{2}{|c|}{Caption-Takeaway 1} & \multicolumn{2}{|c||}{Caption-Takeaway 2} & & \\
 \cline{2-5}
Source & Caption & Takeaway & Caption & Takeaway & $Z$ & $p$\\
 \hline \hline
 \multicolumn{7}{|l|}{Block 1. Takeaways mentioning feature in caption vs. without caption} \\ 
 \hline
\multirow{4}{*}{Synthetic} & 1st & 1st & no cap & 1st & $2.846$ & $0.002^{*}$ \\
 & 2nd & 2nd & no cap & 2nd & $4.641$ & $<0.001^{*}$ \\
 & 3rd & 3rd & no cap & 3rd & $3.643$ & $0.001^{*}$ \\
 & non-pro & non-pro & no cap & non-pro & $6.195$ & $<0.001^{*}$ \\
 \hline
 \multirow{4}{*}{Real-world} & 1st & 1st & no cap & 1st & $1.660$ & $0.049$ \\
 & 2nd & 2nd & no cap & 2nd & $4.225$ & $<0.001^{*}$ \\
 & 3rd & 3rd & no cap & 3rd & $3.347$ & $<0.001^{*}$ \\
 & non-pro & non-pro & no cap & non-pro & $4.732$ & $<0.001^{*}$ \\
 \hline  \hline
 \multicolumn{7}{|l|}{Block 2. Between takeaways mentioning feature in caption} \\  
 \hline
\multirow{3}{*}{Synthetic} & 1st & 1st & 2nd & 2nd & $1.782$ & $0.037$ \\
 & 2nd & 2nd & 3rd & 3rd & $0.705$ & $0.044$ \\
 & 3rd & 3rd & non-pro & non-pro & $8.989$ & $<0.001^{*}$ \\
 \hline
 \multirow{3}{*}{Real-world} & 1st & 1st & 2nd & 2nd & $3.708$ & $<0.001^{*}$ \\
 & 2nd & 2nd & 3rd & 3rd & $0.363$ & $0.358$ \\
 & 3rd & 3rd & non-pro & non-pro & $5.940$ & $<0.001^{*}$ \\
 \hline \hline
 \multicolumn{7}{|l|}{Block 3. When caption $=$ 1st: takeaway $=$ 1st vs. takeaway $\neq$ 1st} \\   
 \hline
\multirow{3}{*}{Synthetic} & 1st & 1st & 1st & 2nd & $8.168$ & $<0.001^{*}$ \\
 & 1st & 1st & 1st & 3rd & $8.275$ & $<0.001^{*}$ \\
 & 1st & 1st & 1st & non-pro & $19.463$ & $<0.001^{*}$ \\
 \hline
 \multirow{3}{*}{Real-world} & 1st & 1st & 1st & 2nd & $9.981$ & $<0.001^{*}$ \\
 & 1st & 1st & 1st & 3rd & $11.301$ & $<0.001^{*}$ \\
 & 1st & 1st & 1st & non-pro & $11.536$ & $<0.001^{*}$ \\
 \hline \hline
 \multicolumn{7}{|l|}{Block 4. When caption $\neq$ 1st: takeaway $=$ 1st vs. takeaway $=$ caption} \\    
 \hline
\multirow{3}{*}{Synthetic} & 2nd & 2nd & 2nd & 1st & $3.829$ & $<0.001^{*}$ \\
 & 3rd & 3rd & 3rd & 1st & $0.258$ & $0.398$ \\
 & non-pro & 1st & non-pro & non-pro & $8.342$ & $<0.001^{*}$ \\
 \hline
 \multirow{3}{*}{Real-world} & 2nd & 2nd & 2nd & 1st & $2.010$ & $0.022$ \\
 & 3rd & 3rd & 3rd & 1st & $2.521$ & $0.006^{*}$ \\
 & non-pro & 1st & non-pro & non-pro & $5.454$ & $<0.001^{*}$ \\
 \hline
\end{tabular}
}
}
\end{center}
\caption{\editmark{Pairwise Z-test results of comparisons between various ratios of takeaways that mention a certain feature (third, fifth columns) when provided a caption describing a certain feature (second, fourth columns). The tests were one-sided with the alternative hypothesis that the ratio of takeaways for `Caption-Takeaway 1' is greater than the ratio of takeaways for `Caption-Takeaway 2'. Asterisks indicate significance with Bonferroni correction.}}
  \label{tab:z-tests}
\end{table}
}
\doublecol{
\begin{table}
\begin{center}
\resizebox{\linewidth}{!}{
\footnotesize{
\begin{tabular}{ |l|l|l|l|l||c|c| }
 \hline
 & \multicolumn{2}{|c|}{Caption-Takeaway 1} & \multicolumn{2}{|c||}{Caption-Takeaway 2} & & \\
 \cline{2-5}
Source & Caption & Takeaway & Caption & Takeaway & $Z$ & $p$\\
 \hline \hline
 \multicolumn{7}{|l|}{Block 1. Takeaways mentioning feature in caption vs. without caption} \\ 
 \hline
\multirow{4}{*}{Synthetic} & 1st & 1st & no cap & 1st & $2.846$ & $0.002^{*}$ \\
 & 2nd & 2nd & no cap & 2nd & $4.641$ & $<0.001^{*}$ \\
 & 3rd & 3rd & no cap & 3rd & $3.643$ & $0.001^{*}$ \\
 & non-pro & non-pro & no cap & non-pro & $6.195$ & $<0.001^{*}$ \\
 \hline
 \multirow{4}{*}{Real-world} & 1st & 1st & no cap & 1st & $1.660$ & $0.049$ \\
 & 2nd & 2nd & no cap & 2nd & $4.225$ & $<0.001^{*}$ \\
 & 3rd & 3rd & no cap & 3rd & $3.347$ & $<0.001^{*}$ \\
 & non-pro & non-pro & no cap & non-pro & $4.732$ & $<0.001^{*}$ \\
 \hline  \hline
 \multicolumn{7}{|l|}{Block 2. Between takeaways mentioning feature in caption} \\  
 \hline
\multirow{3}{*}{Synthetic} & 1st & 1st & 2nd & 2nd & $1.782$ & $0.037$ \\
 & 2nd & 2nd & 3rd & 3rd & $0.705$ & $0.044$ \\
 & 3rd & 3rd & non-pro & non-pro & $8.989$ & $<0.001^{*}$ \\
 \hline
 \multirow{3}{*}{Real-world} & 1st & 1st & 2nd & 2nd & $3.708$ & $<0.001^{*}$ \\
 & 2nd & 2nd & 3rd & 3rd & $0.363$ & $0.358$ \\
 & 3rd & 3rd & non-pro & non-pro & $5.940$ & $<0.001^{*}$ \\
 \hline \hline
 \multicolumn{7}{|l|}{Block 3. When caption $=$ 1st: takeaway $=$ 1st vs. takeaway $\neq$ 1st} \\   
 \hline
\multirow{3}{*}{Synthetic} & 1st & 1st & 1st & 2nd & $8.168$ & $<0.001^{*}$ \\
 & 1st & 1st & 1st & 3rd & $8.275$ & $<0.001^{*}$ \\
 & 1st & 1st & 1st & non-pro & $19.463$ & $<0.001^{*}$ \\
 \hline
 \multirow{3}{*}{Real-world} & 1st & 1st & 1st & 2nd & $9.981$ & $<0.001^{*}$ \\
 & 1st & 1st & 1st & 3rd & $11.301$ & $<0.001^{*}$ \\
 & 1st & 1st & 1st & non-pro & $11.536$ & $<0.001^{*}$ \\
 \hline \hline
 \multicolumn{7}{|l|}{Block 4. When caption $\neq$ 1st: takeaway $=$ 1st vs. takeaway $=$ caption} \\    
 \hline
\multirow{3}{*}{Synthetic} & 2nd & 2nd & 2nd & 1st & $3.829$ & $<0.001^{*}$ \\
 & 3rd & 3rd & 3rd & 1st & $0.258$ & $0.398$ \\
 & non-pro & 1st & non-pro & non-pro & $8.342$ & $<0.001^{*}$ \\
 \hline
 \multirow{3}{*}{Real-world} & 2nd & 2nd & 2nd & 1st & $2.010$ & $0.022$ \\
 & 3rd & 3rd & 3rd & 1st & $2.521$ & $0.006^{*}$ \\
 & non-pro & 1st & non-pro & non-pro & $5.454$ & $<0.001^{*}$ \\
 \hline
\end{tabular}
}
}
\end{center}
\caption{\editmark{Pairwise Z-test results of comparisons between various ratios of takeaways that mention a certain feature (third, fifth columns) when provided a caption describing a certain feature (second, fourth columns). The tests were one-sided with the alternative hypothesis that the ratio of takeaways for `Caption-Takeaway 1' is greater than the ratio of takeaways for `Caption-Takeaway 2'. Asterisks indicate significance with Bonferroni correction.}}
  \label{tab:z-tests}
  \vspace{-5mm}
\end{table}
}

\doublecol{
\begin{figure}
	\centering
    \includegraphics[width=\linewidth]{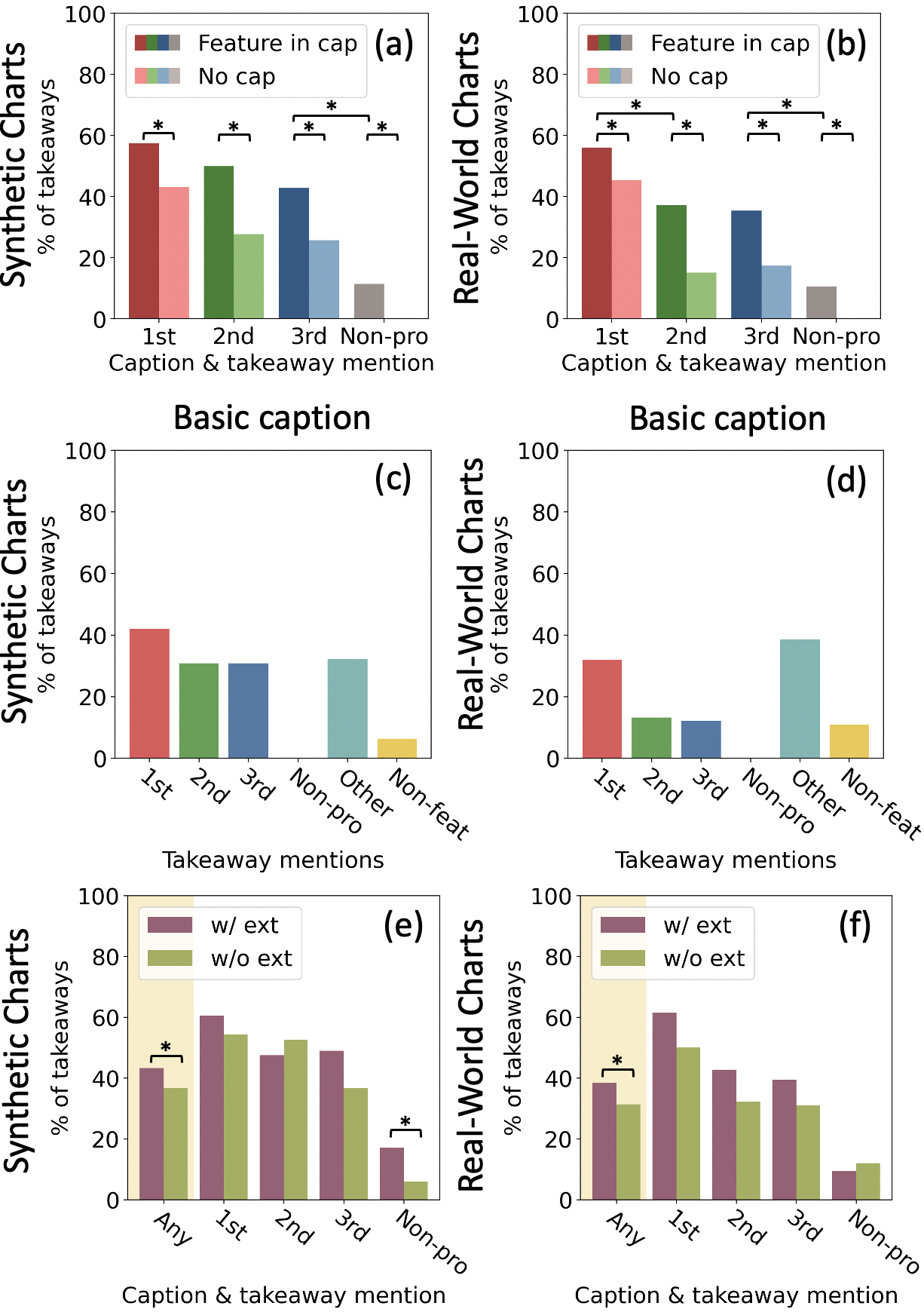}
     \caption{(Top row)
    \editmark{Comparison of percentages of takeaways that mention the same feature as the caption for the synthetic (a) and real-world (b) datasets (i.e., darker bars on the left correspond to the red bar from Figure~\ref{fig:h1-results}a, the green bar from~\ref{fig:h1-results}b, the blue bar from~\ref{fig:h1-results}c, and the grey bar from~\ref{fig:h1-results}d), and percentages of takeaways that mention the feature in the no caption condition (i.e., the right lighter-hued bars in the chart correspond to the bars from Figure~\ref{fig:h1-results}e).} (Middle row) Percentage of takeaways mentioning the visual features at each prominence level when presented with the basic caption. (Bottom row) Dividing the \editmark{left bars in} charts (top row)a and (top row)b based on whether the caption contains external information (purple bars) or does not (olive bars). \editmark{The leftmost \textit{Any} bars show aggregates over all prominence levels. Asterisks indicate significant difference.}}
    \label{fig:collected-result}          
    \Description[(Top row) Comparisons of results in Figure 6. (Middle row) Results with basic caption. (Bottom row) Percentage of takeaways mentioning the visual features at each prominence level when presented with the basic caption.]{(top row) Comparison of percentages of takeaways that mention the same feature as the caption for the synthetic (a) and real-world (b) datasets (i.e., darker bars on the left correspond to the red bar from Figure 6a, the green bar from 6b, the blue bar from 6c, and the grey bar from 6d), and percentages of takeaways that mention the feature in the no caption condition (i.e., the right lighter-hued bars in the chart correspond to the bars from Figure 6e). (middle row) Percentage of takeaways mentioning the visual features at each prominence level when presented with the basic caption. (bottom row) Dividing the left bars in charts (top row)a and (top row)b based on whether the caption contains external information (purple bars) or does not (olive bars). The leftmost Any bars show aggregates over all prominence levels. Asterisks indicate significant difference.}
 \vspace{-4mm}
\end{figure}
}

Figures~\ref{fig:collected-result}a and \ref{fig:collected-result}b collect 
the percentage of takeaways that mention the same feature as in
the caption for the synthetic and the real-world datasets
respectively \editmarkvar{(left darker bars) and compare them with the percentages corresponding to the no-caption case (lighter-hued bars on the right)}. \editmark{We see that captions do play a role in forming takeaways and} the takeaway is \editmark{thus} more likely to mention
that feature (i.e., \editmarkvar{each darker bar in Figures~\ref{fig:collected-result}a and \ref{fig:collected-result}b is usually longer than the corresponding lighter-hued bar to its right}).
\editmarkvar{Planned pairwise Z-tests with Bonferroni correction are shown in Table~\ref{tab:z-tests}. Block 1 shows that the differences between the corresponding color bars are significant for the second most prominent, third most prominent, and non-prominent features. For the most prominent feature, we find that while a higher proportion of people mentioned the most prominent feature in their takeaways when the caption mentions it, the difference is only significant for the synthetic charts. We believe that this is possibly because people already include the most prominent features in their takeaways in the no-caption condition and the difference hence is not significant.} 

\editmarkvar{While we confirmed that both the chart and caption play a role as to what the reader takes away from them, the key question is how the chart and the caption interact with each other -- Do they have a synergistic effect when they emphasize the same feature? Which one wins over when they emphasize different features? Referring to Figure~\ref{fig:h1-results}, we see the synergistic effect of the double-emphasis from the chart and caption when they emphasize the same feature (Figures~\ref{fig:h1-results}a and \ref{fig:h1-results}f). In particular, the participants took away from the most prominent feature significantly more often than from any other feature in the chart (Table~\ref{tab:z-tests} Block 3). 
\cut{Participants took away from the the chart's most prominent feature  when the caption diverged from the most prominent feature and described a non-prominent feature (Figures~\ref{fig:h1-results}d and \ref{fig:h1-results}i, }When the caption diverged from the chart and described a feature that was not prominent, the participants relied more on the chart and took away from the most prominent feature significantly more than the feature described in the caption (Table~\ref{tab:z-tests} Block 4, rows 3 and 6; Figures~\ref{fig:h1-results}d and \ref{fig:h1-results}i). When the caption did not diverge as much and described the second or the third most prominent feature, the takeaways mentioned the feature described in the caption more than the most prominent feature (Table~\ref{tab:z-tests} Block 4, rows 1, 2, 4, and 5; Figures~\ref{fig:h1-results}b, \ref{fig:h1-results}c, \ref{fig:h1-results}g, and \ref{fig:h1-results}h). However, the difference was smaller than the difference between the ratio of people who took away from the most prominent feature and the ratio of people who took away from any of the other features. We believe this result may be due to the fact that the charts still had more influence on the readers than the captions as the second and the third most prominent feature are still among the top prominent features and are among the features emphasized by the chart.}

\singlecol{
\begin{wrapfigure}{r}{0.45\textwidth}
	\centering
    \includegraphics[width=0.45\textwidth]{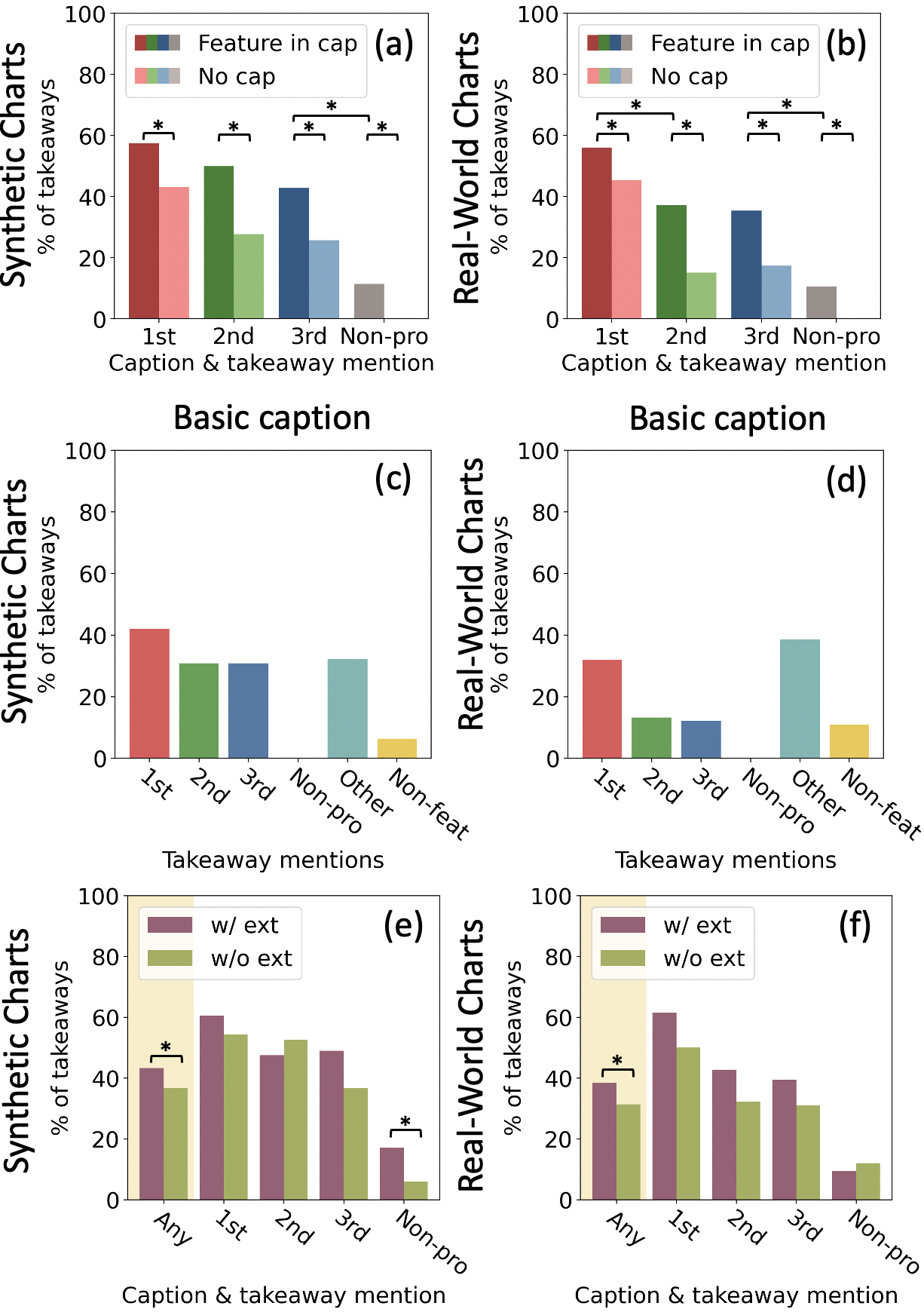}
     \caption{(Top row)
    \editmark{Comparison of percentages of takeaways that mention the same feature as the caption for the synthetic (a) and real-world (b) datasets (i.e., darker bars on the left correspond to the red bar from Figure~\ref{fig:h1-results}a, the green bar from~\ref{fig:h1-results}b, the blue bar from~\ref{fig:h1-results}c, and the grey bar from~\ref{fig:h1-results}d), and percentages of takeaways that mention the feature in the no caption condition (i.e., the right lighter-hued bars in the chart correspond to the bars from Figure~\ref{fig:h1-results}e).} (Middle row) Percentage of takeaways mentioning the visual features at each prominence level when presented with the basic caption. (Bottom row) Dividing the \editmark{left bars in} charts (top row)a and (top row)b based on whether the caption contains external information (purple bars) or does not (olive bars). \editmark{The leftmost \textit{Any} bars show aggregates over all prominence levels. Asterisks indicate significant difference.}}
    \label{fig:collected-result}
          \Description[(Top row) Comparisons of results in Figure 6. (Middle row) Results with basic caption. (Bottom row) Percentage of takeaways mentioning the visual features at each prominence level when presented with the basic caption.]{(top row) Comparison of percentages of takeaways that mention the same feature as the caption for the synthetic (a) and real-world (b) datasets (i.e., darker bars on the left correspond to the red bar from Figure 6a, the green bar from 6b, the blue bar from 6c, and the grey bar from 6d), and percentages of takeaways that mention the feature in the no caption condition (i.e., the right lighter-hued bars in the chart correspond to the bars from Figure 6e). (middle row) Percentage of takeaways mentioning the visual features at each prominence level when presented with the basic caption. (Bottom row) Dividing the left bars in charts (top row)a and (top row)b based on whether the caption contains external information (purple bars) or does not (olive bars). The leftmost Any bars show aggregates over all prominence levels. Asterisks indicate significant difference.}
 \vspace{5mm}
\end{wrapfigure}
}

\editmark{We observe from Figure~\ref{fig:collected-result} that the chart also plays an important role in what people take away -- when a caption mentions a higher-prominent feature, the takeaways more consistently mentions that feature.} Specifically, we see that the bars
for the higher-prominence features are taller than the bars for the
lower-prominence features, indicating an increase in \editmark{the effectiveness of chart in reinforcing the message in the caption.} Planned pairwise Z-tests with Bonferroni correction between each subsequent pair of bars (red bar vs. green bar, green bar vs. blue bar, blue bar vs. gray
bar) (Table~\ref{tab:z-tests} Block 2) find that the red bar vs. green bar is significant \editmark{for real-world charts} and the blue bar
vs. gray bar is significant \editmark{both synthetic and real-world charts}, whereas the green bar
vs. blue bar difference is not significant. We believe that the visual prominence levels for \editmark{some of the top-ranked} features are similar in several charts (i.e., the difference in prominence between the 1st and 2nd ranked features is small) in our dataset and this results in a smaller difference between them, although the trend is in the right direction.

\doublecol{
\begin{table}
\begin{center}
\resizebox{\linewidth}{!}{
\footnotesize{
\begin{tabular}{ |l|l||c|c| }
 \hline
 & & \multicolumn{2}{|c|}{Reported Reliance} \\
 \cline{3-4}
 Source & Caption Type & Chart & Caption \\
 \hline \hline 
 \multicolumn{4}{|l|}{Block 1. Overall} \\   
\hline
 Synthetic & all & $4.675 \pm 0.670$ & $2.624 \pm 1.609$ \\
 \hline
 Real-world & all & $4.536 \pm 0.784$ & $2.779 \pm 1.679$ \\
 \hline \hline
 \multicolumn{4}{|l|}{Block 2. Prominence} \\   
\hline
 \multirow{4}{*}{Synthetic} & 1st & $4.590 \pm 0.711$ & $3.249 \pm 1.327$ \\
 & 2nd & $4.567 \pm 0.814$ & $3.082 \pm 1.433$ \\
 & 3rd & $4.567 \pm 0.726$ & $3.059 \pm 1.408$ \\
 & non-pro & $4.775 \pm 0.549$ & $2.447 \pm 1.429$ \\
 & basic & $4.850 \pm 0.377$ & $2.593 \pm 1.320$ \\
 \hline
 \multirow{4}{*}{Real-world} & 1st & $4.494 \pm 0.838$ & $3.405 \pm 1.481$ \\
 & 2nd & $4.462 \pm 0.890$ & $3.165 \pm 1.359$ \\
 & 3rd & $4.503 \pm 0.805$ & $3.236 \pm 1.354$ \\
 & non-pro & $4.595 \pm 0.718$ & $2.680 \pm 1.545$ \\
 & basic & $4.628 \pm 0.601$ & $2.718 \pm 1.568$ \\
 \hline \hline
 \multicolumn{4}{|l|}{Block 3. External Information} \\   
\hline
 \multirow{2}{*}{Synthetic} & w/o ext & $4.679 \pm 0.688$ & $2.798 \pm 1.402$ \\
 & w/ ext & $4.573 \pm 0.728$ & $3.110 \pm 1.448$ \\
 \hline
 \multirow{2}{*}{Real-world} & w/o ext & $4.606 \pm 0.741$ & $3.061 \pm 1.481$ \\
 & w/ ext & $4.424 \pm 0.875$ & $3.194 \pm 1.439$ \\
 \hline
\end{tabular}
}
}
\end{center}
\caption{\editmark{The reported reliance on the chart and the caption respectively on 5-point Likert scales. Block 1 shows the reported reliance across all the captions. Block 2 shows the reported reliance depending on the prominence of the feature described in the chart and Block 3 shows the reported reliance depending on the inclusion of external information. The values are reported in the form of $\mu \pm \sigma$.}}
  \label{tab:likert-results}
  \vspace{-11.8mm}
\end{table}
}

\editmarkvar{Table~\ref{tab:likert-results} shows average and standard
  deviation of how much the participants reported to have relied on
  the chart and the caption respectively on a 5-point Likert
  scale. 
\editmark{The results in Table~\ref{tab:likert-results} Block 1 suggest that the participants drew information
  from both the chart and the caption when determining their
  takeaways, although they consistently relied on the chart more than
  the caption. These results potentially shed light on why participants took away more often from the chart than the caption when they began to diverge -- they relied more on the chart than the caption.} 
  The results further suggest that the participants' tendency to rely on the
  charts grew while their tendency to rely on the captions declined as
  the prominence of the feature described in the caption
  decreased (Table~\ref{tab:likert-results} Block 2). We found a
  significant drop in the self-reported reliance on the caption when the
  caption described a non-prominent feature compared to when it
  described the third-most prominent feature (synthetic: Mann-Whitney
  $U = 28941$, $p < 0.001$; real-world: Mann-Whitney $U = 9666$, $p <
  0.001$) whereas the increase in the reported reliance on the chart
  when the
  caption described a non-prominent feature compared to when it
  described the third-most prominent feature was only significant with the synthetic charts
  (Mann-Whitney $U = 32844.5$, $p < 0.001$). Although the general
  trend is in the right direction, we did not find significant
  differences in the reliance scores when the caption mentioned one of
  the top three prominent features. This may be because the difference in
  prominence is not as great among these features as it is with the
  non-prominent feature. These results are in line with our findings from the takeaways; we find that when the chart contains a high-prominence visual feature, but the caption emphasizes a low-prominence feature, participants relied more on the chart and less on the caption.}

\singlecol{
\begin{table}
\begin{center}
\resizebox{0.6\linewidth}{!}{
\footnotesize{
\begin{tabular}{ |l|l||c|c| }
 \hline
 & & \multicolumn{2}{|c|}{Reported Reliance} \\
 \cline{3-4}
 Source & Caption Type & Chart & Caption \\
 \hline \hline 
 \multicolumn{4}{|l|}{Block 1. Overall} \\   
\hline
 Synthetic & all & $4.675 \pm 0.670$ & $2.624 \pm 1.609$ \\
 \hline
 Real-world & all & $4.536 \pm 0.784$ & $2.779 \pm 1.679$ \\
 \hline \hline
 \multicolumn{4}{|l|}{Block 2. Prominence} \\   
\hline
 \multirow{4}{*}{Synthetic} & 1st & $4.590 \pm 0.711$ & $3.249 \pm 1.327$ \\
 & 2nd & $4.567 \pm 0.814$ & $3.082 \pm 1.433$ \\
 & 3rd & $4.567 \pm 0.726$ & $3.059 \pm 1.408$ \\
 & non-pro & $4.775 \pm 0.549$ & $2.447 \pm 1.429$ \\
 & basic & $4.850 \pm 0.377$ & $2.593 \pm 1.320$ \\
 \hline
 \multirow{4}{*}{Real-world} & 1st & $4.494 \pm 0.838$ & $3.405 \pm 1.481$ \\
 & 2nd & $4.462 \pm 0.890$ & $3.165 \pm 1.359$ \\
 & 3rd & $4.503 \pm 0.805$ & $3.236 \pm 1.354$ \\
 & non-pro & $4.595 \pm 0.718$ & $2.680 \pm 1.545$ \\
 & basic & $4.628 \pm 0.601$ & $2.718 \pm 1.568$ \\
 \hline \hline
 \multicolumn{4}{|l|}{Block 3. External Information} \\   
\hline
 \multirow{2}{*}{Synthetic} & w/o ext & $4.679 \pm 0.688$ & $2.798 \pm 1.402$ \\
 & w/ ext & $4.573 \pm 0.728$ & $3.110 \pm 1.448$ \\
 \hline
 \multirow{2}{*}{Real-world} & w/o ext & $4.606 \pm 0.741$ & $3.061 \pm 1.481$ \\
 & w/ ext & $4.424 \pm 0.875$ & $3.194 \pm 1.439$ \\
 \hline
\end{tabular}
}
}
\end{center}
\caption{\editmark{The reported reliance on the chart and the caption respectively on 5-point Likert scales. Block 1 shows the reported reliance across all the captions. Block 2 shows the reported reliance depending on the prominence of the feature described in the chart and Block 3 shows the reported reliance depending on the inclusion of external information. The values are reported in the form of $\mu \pm \sigma$.}}
  \label{tab:likert-results}
\end{table}
}

Considering all these results together suggests that we can accept our
hypothesis {\bf H1} \editmark{ -- readers take away from the highly
  prominent features when the chart and caption both emphasize the
  same feature and that their inclination to rely more on the most
  prominent feature instead of the feature described in the caption
  becomes greater when the caption describes a less prominent
  feature.}

\vspace{0.05in}
\noindent \editmark{{\bf \em H1 Additional Results.}}
We also collected takeaways for charts with {\em basic} captions that
describe the axes of the chart.  (Figure~\ref{fig:collected-result}
- middle row). We find that the percentage of takeaways for each of
the features is similar to that of the no-caption condition. In fact,
Pearson's chi-square test finds no significant difference between the
takeaway histograms of the basic caption and the no-caption conditions
(synthetic: \editmark{$\chi^{2} (4) = 1.564$}, \editmark{$p = 0.815$};
real-world: \editmark{$\chi^{2} (4) = 7.168$}, \editmark{$p =
  0.127$}). \editmark{While automated captioning tools~\cite{tableau,powerbi} generate captions corresponding to our basic captions, we were unable to find evidence that these captions affect what people take away. Such captions may help readers with accessibility needs; however, we believe further exploration will help future systems determine appropriate uses for such captions.}

\vspace{0.05in}
\noindent {\bf \em Assessing H2.}  To evaluate {\bf H2}, we examine
whether including external content information in the caption makes it
more likely for readers to take away the feature mentioned in the
caption. \editmark{We find that
  people are significantly more likely to mention the feature
  described in the caption when it includes external information than when it does not (Figures~\ref{fig:collected-result}e and Figures~\ref{fig:collected-result}f \textit{Any} bars). A pairwise Z-test finds significant difference between these ratios (synthetic: $Z =
  2.273, p = 0.011$; real-world: $Z = 2.032, p = 0.021$)}.
\editmarkvar{In addition, the reported reliance on the chart and the captions shifted towards the captions with external information, which is in-line with our findings (Figure~\ref{tab:likert-results} Block 3). Specifically, the reported reliance on the chart was significantly lower with external information (synthetic: Mann-Whitney $U =  137318$, $p < 0.001$; real-world: Mann-Whitney $U = 45292$, $p = 0.001$); the reported reliance on the caption was higher with external information, but the difference was only significant for the synthetic charts (synthetic: Mann-Whitney $U =  131594$, $p < 0.001$; real-world: Mann-Whitney $U = 48599.5$, $p = 0.132$).}

\editmarkvar{The results together suggest that we can accept {\bf H2} that states that including external information in the caption helps reinforce the message in the caption and users are more likely to take away from the feature described in the caption.}

\vspace{0.05in}
\noindent \editmark{{\bf \em H2 Additional Results.}}
\editmark{Figure~\ref{fig:collected-result} (bottom row) breaks down the ratio of the takeaways that mention the feature described in the caption by level of prominence of the feature.
The figure shows that there is usually an increase in the ratio of the takeaways that mentioned the feature described in the caption when the caption included external information for each level of prominence. Among the differences, we only found significant difference when the caption mentioned a non-prominent feature for synthetic charts ($Z = 3.027$, $p = 0.001$). Further study could shed light on the correlation between the prominence of the feature described in the caption and how external information affects the readers' takeaways.} 

\section{Design Guidelines}

\begin{figure}[t]
\centering
\subfigure[``The cheap Yen and PM Abe's tourism policy caused the number of tourists in Japan to steeply rise between 2011 and 2018."]{\includegraphics[width=0.48\textwidth]{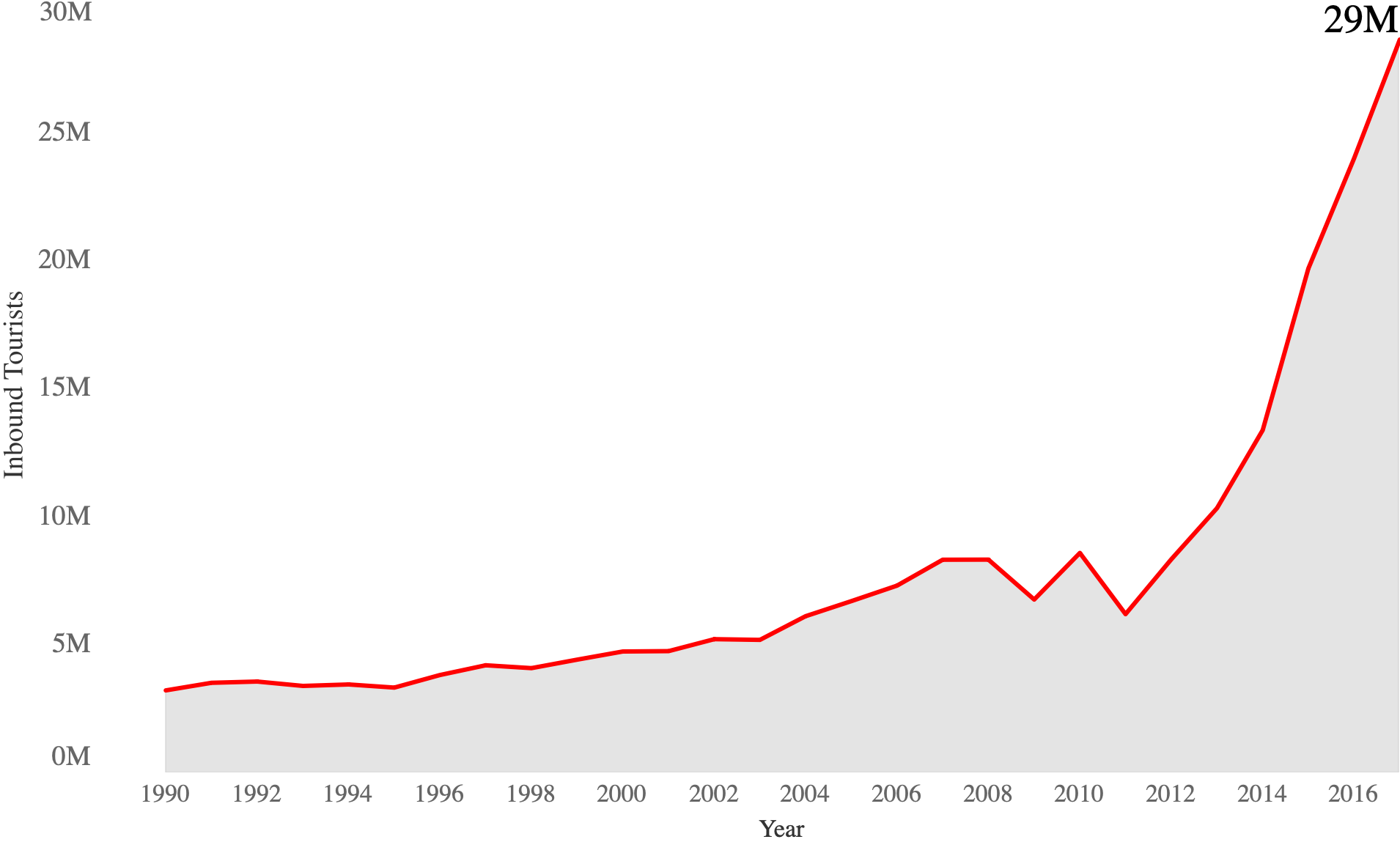}}
\subfigure[``Due to the 2008 Financial Crisis, the number of tourists in Japan decreased in 2009."]{\includegraphics[width=0.48\textwidth]{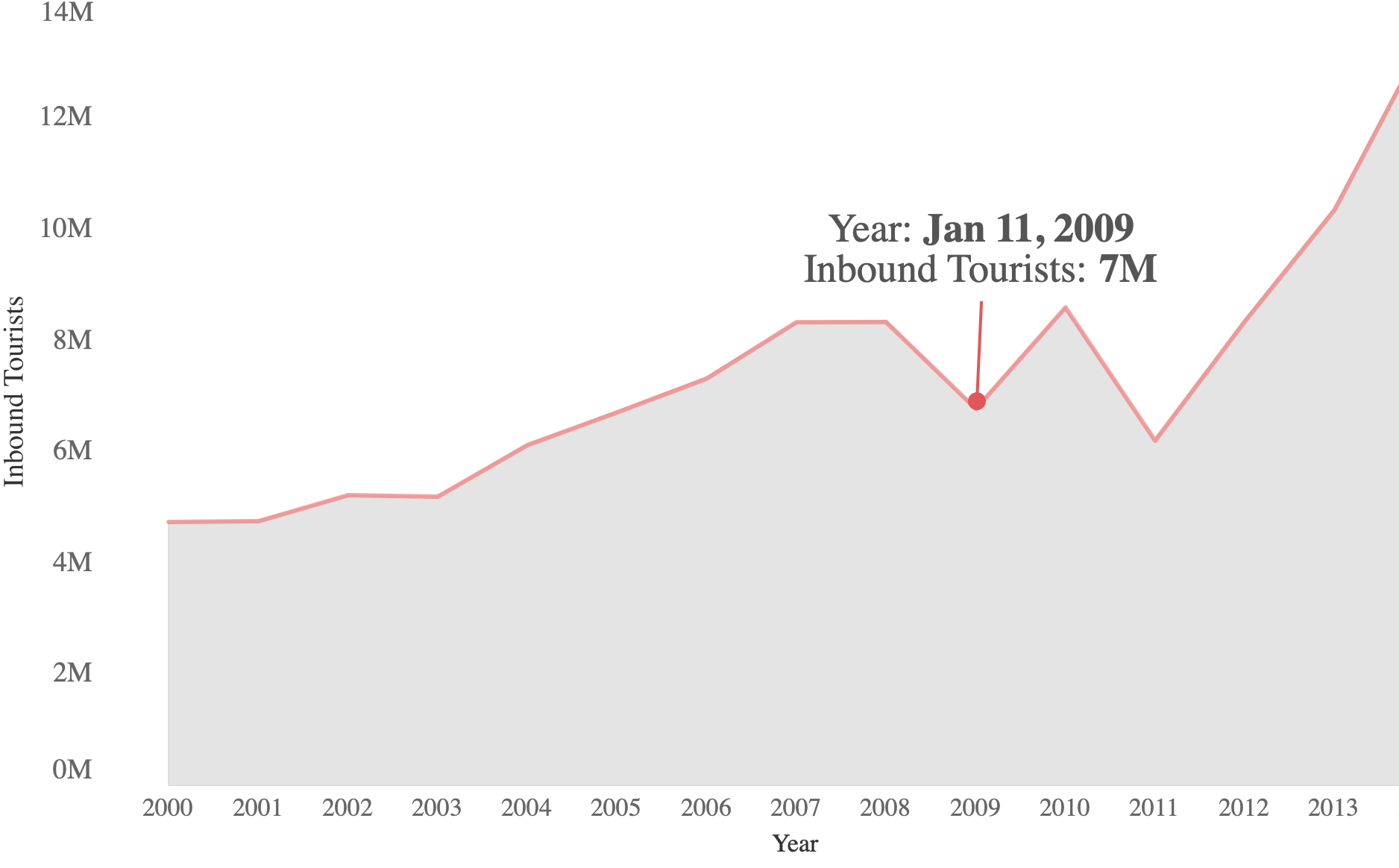}}
\vspace{-5mm}
\caption{Examples of chart-caption pairs authored to emphasize the same feature in the data. (a) Both the caption and chart emphasize the sharp positive trend. (b) The original chart is modified to zoom into a portion of the time range and the feature is made more visually prominent with an annotation showing the dip in the number of tourists. The caption describes that dip with additional context.}
\Description[Examples of chart-caption pairs authored to emphasize the same feature in the data together.]{Examples of chart-caption pairs authored to emphasize the same feature in the data. (a) Both the caption and chart emphasize the sharp positive trend. (b) The original chart is modified to zoom into a portion of the time range and the feature is made more visually prominent with an annotation showing the dip in the number of tourists. The caption describes that dip with additional context.}
\label{fig:guidelines}
\vspace{-5mm}
\end{figure}

\editmark{Our findings indicate that the readers will take away from the feature doubly emphasized by both the chart and caption if they provide a coherent message. However, when the chart and caption diverge in terms of the feature that they are emphasizing, readers are less likely to use information from the caption in their takeaways. To improve the efficacy of the chart-caption pair, authors could (1) design the chart to make the feature described in the caption more prominent and (2) include external information in the caption to give more context to the information in the caption.}

There are several ways for authors to emphasize aspects of the data in a chart so that readers' attention is drawn to these visual features.  \editmark{One technique is to ensure that aspects of the data such as trends and outliers are presented at the right level of detail or interval range; too-broad of a measurement interval may hide a signal. For example, assume that we were given the chart in Figure~\ref{fig:guidelines}a with the caption in Figure~\ref{fig:guidelines}b. The decrease in 2009 is not very prominent because the large increase starting in 2011 overshadows the decrease. Zooming closer to the intended feature and cropping out irrelevant features (Figure~\ref{fig:guidelines}b), helps make the feature more visually prominent.}
\editmark{However, when zooming into the data in this manner, authors must take precaution to avoid removing important information or rendering the chart misleading~\cite{o2018testing, pandey2015deceptive}.}

A simple way to \editmark{further} facilitate effective chart reading
is to enhance the visualization with highlighting and overlays such as annotations to guide the audience's attention to the
image area they are describing~\cite{kong:2012}
\editmark{(Figure~\ref{fig:guidelines}b)}. Sometimes, a different
chart altogether may be more effective to emphasize a particular
aspect of the data. For example, converting continuous data in line
charts into discrete values could help emphasize individual values
that the author would like to focus on. The consistency between the
redesigned chart-caption pairs helps readers take away from the doubly
emphasized feature (Figure~\ref{fig:guidelines}).


\section{Future Work}
\noindent \editmark{{\bf \em Chart and caption authoring tools.}}
We would like to explore how this work can provide interesting implications for \emph{both} chart and caption design to help the author effectively convey a specific point of view. Enhancements to visualization authoring tools could suggest chart design alternatives given a feature that the author would like to emphasize. 
Specifically, the system could go further by emphasizing features in the chart according to the main message the author wants to convey by automatically adding annotations to the chart, adding
highlights, and adjusting levels of detail so that the chart and the caption deliver a concerted message. This will require formulating a high-level language specification that the authors can use to communicate to the system about their intents or a natural language processing module that can infer the authors' intents based on the captions they write. 
Coordinating interaction between the chart and the caption such that hovering over the text in the caption would highlight the corresponding visual feature in the chart and vice-versa, is another interesting direction to pursue to help the reader. \editmark{The resulting system would be a significant extension of the \textit{interactive document reader} presented by Kong et al. and Kim et al.~\cite{Kong2014extracting, Kim2018facilitating}.} 
\editmark{On the captioning side, a system could classify basic captions, captions about high-prominence features, and captions about low-prominence features. Based on the classification, the system could suggest external information to further emphasize the information presented.}

\vspace{0.05in}
\noindent \editmark{{\bf \em Further exploration of caption variations.} In this work, we use a template-based approach for
  generating captions to minimize the effect of the variation of
  natural language and to keep the experiment size reasonable.
  Simultaneously, we carefully vary the visual feature described in
  the caption and the presence of external information to best
  understand how people read captions and charts together to form
  their takeaways. Future work could study captions with various
  natural language expressions and different ways of emphasis. It would be useful to understand whether the relationship between multiple features in a caption (e.g., a simple list - \textit{``There were major dips in employment in 2008 and 2020.''} or a comparison - \textit{``The dip in 2020 was greater than the dip in 2008.''}) has an effect on what readers take away. Studying how our findings generalize to other types of external information (e.g., extrapolation, breakdown into subcategories) would be an interesting direction  to pursue.}

\vspace{0.05in}
\noindent \editmark{{\bf \em Generalization to other chart types.}} Our work
explores how readers take away information when presented with
univariate line charts and captions. \editmark{Basic chart types still have prominent features (e.g., extrema in bar charts, outliers in scatterplots) and less prominent features (e.g., a point in a cluster in scatterplots). We expect similar findings would hold for those other chart types. We leave it to future work to confirm this intuition.}


\section{Conclusion}
In this paper, we examine what readers take away from both a chart and its caption. Our results suggest that when the caption mentions visual
features of differing prominence levels, the takeaways differ. When the caption mentions a specific feature, the takeaways also tend to mention that feature.
We also observed that when a caption mentions a \editmark{visually prominent} feature, the takeaways more consistently mention that feature. On the other hand, when the caption mentions \editmark{a less prominent feature},
\editmark{the readers' takeaways are more likely to mention the most prominent prominence features than the feature described in the caption. We also find that including external information in the caption makes the readers more likely to form their takeaways based on the feature described in the caption.}
From the results of our study, we propose guidelines to better design charts and captions together; using visual cues and alternative chart representations, visual features can be made more prominent and be further emphasized by their descriptions in the caption. Design implications from this work provide opportunities for the authoring of chart and caption pairs in visual analysis tools to effectively convey a specific point of view to the reader.

\vspace{-1.1mm}
\begin{acks}
The authors thank the Stanford University HCI Group and Tableau Research for their feedback during the development of the studies. The authors also thank the reviewers for their feedback during the review cycle. This work is supported by NSF award III-1714647.
\end{acks}

\newpage

\bibliographystyle{ACM-Reference-Format}
\bibliography{bibliography}


\begin{thebibliography}{48}


\ifx \showCODEN    \undefined \def \showCODEN     #1{\unskip}     \fi
\ifx \showDOI      \undefined \def \showDOI       #1{#1}\fi
\ifx \showISBNx    \undefined \def \showISBNx     #1{\unskip}     \fi
\ifx \showISBNxiii \undefined \def \showISBNxiii  #1{\unskip}     \fi
\ifx \showISSN     \undefined \def \showISSN      #1{\unskip}     \fi
\ifx \showLCCN     \undefined \def \showLCCN      #1{\unskip}     \fi
\ifx \shownote     \undefined \def \shownote      #1{#1}          \fi
\ifx \showarticletitle \undefined \def \showarticletitle #1{#1}   \fi
\ifx \showURL      \undefined \def \showURL       {\relax}        \fi
\providecommand\bibfield[2]{#2}
\providecommand\bibinfo[2]{#2}
\providecommand\natexlab[1]{#1}
\providecommand\showeprint[2][]{arXiv:#2}

\bibitem[\protect\citeauthoryear{{A}mazon {M}echanical {T}urk}{{A}mazon
  {M}echanical {T}urk}{2020}]%
        {turk}
{A}mazon {M}echanical {T}urk \bibinfo{year}{2020}\natexlab{}.
\newblock \bibinfo{title}{{A}mazon {M}echanical {T}urk}.
\newblock \bibinfo{howpublished}{\url{https://www.mturk.com}}.
\newblock


\bibitem[\protect\citeauthoryear{Borkin, Vo, Bylinskii, Isola, Sunkavalli,
  Oliva, and Pfister}{Borkin et~al\mbox{.}}{2013}]%
        {borkin2013makes}
\bibfield{author}{\bibinfo{person}{Michelle~A Borkin},
  \bibinfo{person}{Azalea~A Vo}, \bibinfo{person}{Zoya Bylinskii},
  \bibinfo{person}{Phillip Isola}, \bibinfo{person}{Shashank Sunkavalli},
  \bibinfo{person}{Aude Oliva}, {and} \bibinfo{person}{Hanspeter Pfister}.}
  \bibinfo{year}{2013}\natexlab{}.
\newblock \showarticletitle{What makes a visualization memorable?}
\newblock \bibinfo{journal}{\emph{IEEE Transactions on Visualization and
  Computer Graphics}} \bibinfo{volume}{19}, \bibinfo{number}{12}
  (\bibinfo{year}{2013}), \bibinfo{pages}{2306--2315}.
\newblock


\bibitem[\protect\citeauthoryear{Bransford}{Bransford}{1979}]%
        {bransford1979human}
\bibfield{author}{\bibinfo{person}{J. Bransford}.}
  \bibinfo{year}{1979}\natexlab{}.
\newblock \bibinfo{booktitle}{\emph{Human Cognition: Learning, Understanding,
  and Remembering}}.
\newblock \bibinfo{publisher}{Wadsworth Publishing Company}.
\newblock
\showISBNx{9780534006990}
\showLCCN{lc78031698}
\urldef\tempurl%
\url{https://books.google.com/books?id=7TKLnQEACAAJ}
\showURL{%
\tempurl}


\bibitem[\protect\citeauthoryear{Carberry, Elzer, and Demir}{Carberry
  et~al\mbox{.}}{2006}]%
        {carberry:2006}
\bibfield{author}{\bibinfo{person}{Sandra Carberry}, \bibinfo{person}{Stephanie
  Elzer}, {and} \bibinfo{person}{Seniz Demir}.}
  \bibinfo{year}{2006}\natexlab{}.
\newblock \showarticletitle{Information Graphics: An Untapped Resource for
  Digital Libraries}. In \bibinfo{booktitle}{\emph{Proceedings of the 29th
  Annual International ACM SIGIR Conference on Research and Development in
  Information Retrieval}} (Seattle, Washington, USA)
  \emph{(\bibinfo{series}{SIGIR '06})}. \bibinfo{publisher}{Association for
  Computing Machinery}, \bibinfo{address}{New York, NY, USA},
  \bibinfo{pages}{581–588}.
\newblock
\showISBNx{1595933697}
\urldef\tempurl%
\url{https://doi.org/10.1145/1148170.1148270}
\showDOI{\tempurl}


\bibitem[\protect\citeauthoryear{Card, Mackinlay, and Shneiderman}{Card
  et~al\mbox{.}}{1999}]%
        {card:1999}
\bibfield{editor}{\bibinfo{person}{Stuart~K. Card}, \bibinfo{person}{Jock~D.
  Mackinlay}, {and} \bibinfo{person}{Ben Shneiderman}} (Eds.).
  \bibinfo{year}{1999}\natexlab{}.
\newblock \bibinfo{booktitle}{\emph{Readings in Information Visualization:
  Using Vision to Think}}.
\newblock \bibinfo{publisher}{Morgan Kaufmann Publishers Inc.},
  \bibinfo{address}{San Francisco, CA, USA}.
\newblock
\showISBNx{1558605339}


\bibitem[\protect\citeauthoryear{Chen, Zhang, Kim, Cohen, Yu, Rossi, and
  Bunescu}{Chen et~al\mbox{.}}{2019a}]%
        {chen2019neural}
\bibfield{author}{\bibinfo{person}{Charles Chen}, \bibinfo{person}{Ruiyi
  Zhang}, \bibinfo{person}{Sungchul Kim}, \bibinfo{person}{Scott Cohen},
  \bibinfo{person}{Tong Yu}, \bibinfo{person}{Ryan Rossi}, {and}
  \bibinfo{person}{Razvan Bunescu}.} \bibinfo{year}{2019}\natexlab{a}.
\newblock \showarticletitle{Neural caption generation over figures}. In
  \bibinfo{booktitle}{\emph{Adjunct Proceedings of the 2019 ACM International
  Joint Conference on Pervasive and Ubiquitous Computing and Proceedings of the
  2019 ACM International Symposium on Wearable Computers}}.
  \bibinfo{pages}{482--485}.
\newblock


\bibitem[\protect\citeauthoryear{Chen, Zhang, Koh, Kim, Cohen, Yu, Rossi, and
  Bunescu}{Chen et~al\mbox{.}}{2019b}]%
        {chen2019figure}
\bibfield{author}{\bibinfo{person}{Charles Chen}, \bibinfo{person}{Ruiyi
  Zhang}, \bibinfo{person}{Eunyee Koh}, \bibinfo{person}{Sungchul Kim},
  \bibinfo{person}{Scott Cohen}, \bibinfo{person}{Tong Yu},
  \bibinfo{person}{Ryan Rossi}, {and} \bibinfo{person}{Razvan Bunescu}.}
  \bibinfo{year}{2019}\natexlab{b}.
\newblock \showarticletitle{Figure captioning with reasoning and sequence-level
  training}.
\newblock \bibinfo{journal}{\emph{arXiv preprint arXiv:1906.02850}}
  (\bibinfo{year}{2019}).
\newblock


\bibitem[\protect\citeauthoryear{Cui, Badam, Yal{\c{c}}in, and Elmqvist}{Cui
  et~al\mbox{.}}{2018}]%
        {datasite}
\bibfield{author}{\bibinfo{person}{Zhe Cui}, \bibinfo{person}{Sriram~Karthik
  Badam}, \bibinfo{person}{Adil Yal{\c{c}}in}, {and} \bibinfo{person}{Niklas
  Elmqvist}.} \bibinfo{year}{2018}\natexlab{}.
\newblock \showarticletitle{DataSite: Proactive Visual Data Exploration with
  Computation of Insight-based Recommendations}.
\newblock \bibinfo{journal}{\emph{CoRR}}  \bibinfo{volume}{abs/1802.08621}
  (\bibinfo{year}{2018}).
\newblock
\showeprint[arxiv]{1802.08621}
\urldef\tempurl%
\url{http://arxiv.org/abs/1802.08621}
\showURL{%
\tempurl}


\bibitem[\protect\citeauthoryear{Demiralp, Haas, Parthasarathy, and
  Pedapati}{Demiralp et~al\mbox{.}}{2017}]%
        {demiralp2017foresight}
\bibfield{author}{\bibinfo{person}{{\c{C}}a{\u{g}}atay Demiralp},
  \bibinfo{person}{Peter~J Haas}, \bibinfo{person}{Srinivasan Parthasarathy},
  {and} \bibinfo{person}{Tejaswini Pedapati}.} \bibinfo{year}{2017}\natexlab{}.
\newblock \showarticletitle{Foresight: Rapid data exploration through
  guideposts}.
\newblock \bibinfo{journal}{\emph{arXiv preprint arXiv:1709.10513}}
  (\bibinfo{year}{2017}).
\newblock


\bibitem[\protect\citeauthoryear{Egeth and Yantis}{Egeth and Yantis}{1997}]%
        {Egeth1997VisualAC}
\bibfield{author}{\bibinfo{person}{H. Egeth} {and} \bibinfo{person}{S.
  Yantis}.} \bibinfo{year}{1997}\natexlab{}.
\newblock \showarticletitle{Visual attention: control, representation, and time
  course.}
\newblock \bibinfo{journal}{\emph{Annual review of psychology}}
  \bibinfo{volume}{48} (\bibinfo{year}{1997}), \bibinfo{pages}{269--97}.
\newblock


\bibitem[\protect\citeauthoryear{Elzer, Carberry, Chester, Demir, Green,
  Zukerman, and Trnka}{Elzer et~al\mbox{.}}{2005}]%
        {elzer2005exploring}
\bibfield{author}{\bibinfo{person}{Stephanie Elzer}, \bibinfo{person}{Sandra
  Carberry}, \bibinfo{person}{Daniel Chester}, \bibinfo{person}{Seniz Demir},
  \bibinfo{person}{Nancy Green}, \bibinfo{person}{Ingrid Zukerman}, {and}
  \bibinfo{person}{Keith Trnka}.} \bibinfo{year}{2005}\natexlab{}.
\newblock \showarticletitle{Exploring and exploiting the limited utility of
  captions in recognizing intention in information graphics}. In
  \bibinfo{booktitle}{\emph{ACL}}. \bibinfo{pages}{223--230}.
\newblock


\bibitem[\protect\citeauthoryear{Elzer, Carberry, and Zukerman}{Elzer
  et~al\mbox{.}}{2011}]%
        {elzer2011automated}
\bibfield{author}{\bibinfo{person}{Stephanie Elzer}, \bibinfo{person}{Sandra
  Carberry}, {and} \bibinfo{person}{Ingrid Zukerman}.}
  \bibinfo{year}{2011}\natexlab{}.
\newblock \showarticletitle{The automated understanding of simple bar charts}.
\newblock \bibinfo{journal}{\emph{Artificial Intelligence}}
  \bibinfo{volume}{175}, \bibinfo{number}{2} (\bibinfo{year}{2011}),
  \bibinfo{pages}{526--555}.
\newblock


\bibitem[\protect\citeauthoryear{Fasciano and Lapalme}{Fasciano and
  Lapalme}{1996}]%
        {fasciano1996postgraphe}
\bibfield{author}{\bibinfo{person}{Massimo Fasciano} {and} \bibinfo{person}{Guy
  Lapalme}.} \bibinfo{year}{1996}\natexlab{}.
\newblock \showarticletitle{Postgraphe: a system for the generation of
  statistical graphics and text}. In \bibinfo{booktitle}{\emph{Eighth
  International Natural Language Generation Workshop}}.
\newblock


\bibitem[\protect\citeauthoryear{Ferres, Verkhogliad, Lindgaard, Boucher,
  Chretien, and Lachance}{Ferres et~al\mbox{.}}{2007}]%
        {ferres:2007}
\bibfield{author}{\bibinfo{person}{Leo Ferres}, \bibinfo{person}{Petro
  Verkhogliad}, \bibinfo{person}{Gitte Lindgaard}, \bibinfo{person}{Louis
  Boucher}, \bibinfo{person}{Antoine Chretien}, {and} \bibinfo{person}{Martin
  Lachance}.} \bibinfo{year}{2007}\natexlab{}.
\newblock \showarticletitle{Improving Accessibility to Statistical Graphs: The
  IGraph-Lite System}. In \bibinfo{booktitle}{\emph{Proceedings of the 9th
  International ACM SIGACCESS Conference on Computers and Accessibility}}
  (Tempe, Arizona, USA) \emph{(\bibinfo{series}{Assets '07})}.
  \bibinfo{publisher}{Association for Computing Machinery},
  \bibinfo{address}{New York, NY, USA}, \bibinfo{pages}{67--74}.
\newblock
\showISBNx{9781595935731}
\urldef\tempurl%
\url{https://doi.org/10.1145/1296843.1296857}
\showDOI{\tempurl}


\bibitem[\protect\citeauthoryear{Hegarty and Just}{Hegarty and Just}{1993}]%
        {hegarty:1993}
\bibfield{author}{\bibinfo{person}{Mary Hegarty} {and}
  \bibinfo{person}{Marcel-Adam Just}.} \bibinfo{year}{1993}\natexlab{}.
\newblock \showarticletitle{Constructing mental models of machines from text
  and diagrams}.
\newblock \bibinfo{journal}{\emph{Journal of memory and language}}
  \bibinfo{volume}{32}, \bibinfo{number}{6} (\bibinfo{year}{1993}),
  \bibinfo{pages}{717--742}.
\newblock


\bibitem[\protect\citeauthoryear{Hu, Orghian, and Hidalgo}{Hu
  et~al\mbox{.}}{2018}]%
        {hu2018dive}
\bibfield{author}{\bibinfo{person}{Kevin Hu}, \bibinfo{person}{Diana Orghian},
  {and} \bibinfo{person}{C{\'e}sar Hidalgo}.} \bibinfo{year}{2018}\natexlab{}.
\newblock \showarticletitle{DIVE: A Mixed-Initiative System Supporting
  Integrated Data Exploration Workflows}. In
  \bibinfo{booktitle}{\emph{Proceedings of the Workshop on Human-In-the-Loop
  Data Analytics}}. ACM, \bibinfo{pages}{5}.
\newblock


\bibitem[\protect\citeauthoryear{Hullman, Diakopoulos, and Adar}{Hullman
  et~al\mbox{.}}{2013}]%
        {hullman2013contextifier}
\bibfield{author}{\bibinfo{person}{Jessica Hullman}, \bibinfo{person}{Nicholas
  Diakopoulos}, {and} \bibinfo{person}{Eytan Adar}.}
  \bibinfo{year}{2013}\natexlab{}.
\newblock \showarticletitle{Contextifier: automatic generation of annotated
  stock visualizations}. In \bibinfo{booktitle}{\emph{Proceedings of the SIGCHI
  Conference on Human Factors in Computing Systems}}.
  \bibinfo{pages}{2707--2716}.
\newblock


\bibitem[\protect\citeauthoryear{Kim, Hoque, and Agrawala}{Kim
  et~al\mbox{.}}{2020}]%
        {kim2020answering}
\bibfield{author}{\bibinfo{person}{Dae~Hyun Kim}, \bibinfo{person}{Enamul
  Hoque}, {and} \bibinfo{person}{Maneesh Agrawala}.}
  \bibinfo{year}{2020}\natexlab{}.
\newblock \showarticletitle{Answering Questions about Charts and Generating
  Visual Explanations}. In \bibinfo{booktitle}{\emph{Proceedings of the 2020
  CHI Conference on Human Factors in Computing Systems}}.
  \bibinfo{pages}{1--13}.
\newblock


\bibitem[\protect\citeauthoryear{Kim, Hoque, Kim, and Agrawala}{Kim
  et~al\mbox{.}}{2018}]%
        {Kim2018facilitating}
\bibfield{author}{\bibinfo{person}{Dae~Hyun Kim}, \bibinfo{person}{Enamul
  Hoque}, \bibinfo{person}{Juho Kim}, {and} \bibinfo{person}{Maneesh
  Agrawala}.} \bibinfo{year}{2018}\natexlab{}.
\newblock \showarticletitle{Facilitating Document Reading by Linking Text and
  Tables}. In \bibinfo{booktitle}{\emph{UIST}} \emph{(\bibinfo{series}{UIST
  '18})}. \bibinfo{publisher}{ACM}, \bibinfo{pages}{423--434}.
\newblock


\bibitem[\protect\citeauthoryear{Kong, Liu, and Karahalios}{Kong
  et~al\mbox{.}}{2018}]%
        {Kong2018FramesAS}
\bibfield{author}{\bibinfo{person}{Ha-Kyung Kong}, \bibinfo{person}{Zhicheng
  Liu}, {and} \bibinfo{person}{Karrie Karahalios}.}
  \bibinfo{year}{2018}\natexlab{}.
\newblock \showarticletitle{Frames and slants in titles of visualizations on
  controversial topics}. In \bibinfo{booktitle}{\emph{Proceedings of the 2018
  CHI Conference on Human Factors in Computing Systems}}.
  \bibinfo{pages}{1--12}.
\newblock


\bibitem[\protect\citeauthoryear{Kong, Liu, and Karahalios}{Kong
  et~al\mbox{.}}{2019}]%
        {kong:2019}
\bibfield{author}{\bibinfo{person}{Ha-Kyung Kong}, \bibinfo{person}{Zhicheng
  Liu}, {and} \bibinfo{person}{Karrie Karahalios}.}
  \bibinfo{year}{2019}\natexlab{}.
\newblock \showarticletitle{Trust and Recall of Information across Varying
  Degrees of Title-Visualization Misalignment}. In
  \bibinfo{booktitle}{\emph{Proceedings of the 2019 CHI Conference on Human
  Factors in Computing Systems}} (Glasgow, Scotland Uk)
  \emph{(\bibinfo{series}{CHI '19})}. \bibinfo{publisher}{Association for
  Computing Machinery}, \bibinfo{address}{New York, NY, USA},
  \bibinfo{pages}{1--13}.
\newblock
\showISBNx{9781450359702}
\urldef\tempurl%
\url{https://doi.org/10.1145/3290605.3300576}
\showDOI{\tempurl}


\bibitem[\protect\citeauthoryear{{Kong} and {Agrawala}}{{Kong} and
  {Agrawala}}{2012}]%
        {kong:2012}
\bibfield{author}{\bibinfo{person}{N. {Kong}} {and} \bibinfo{person}{M.
  {Agrawala}}.} \bibinfo{year}{2012}\natexlab{}.
\newblock \showarticletitle{Graphical Overlays: Using Layered Elements to Aid
  Chart Reading}.
\newblock \bibinfo{journal}{\emph{IEEE Transactions on Visualization and
  Computer Graphics}} \bibinfo{volume}{18}, \bibinfo{number}{12}
  (\bibinfo{year}{2012}), \bibinfo{pages}{2631--2638}.
\newblock


\bibitem[\protect\citeauthoryear{Kong, Hearst, and Agrawala}{Kong
  et~al\mbox{.}}{2014}]%
        {Kong2014extracting}
\bibfield{author}{\bibinfo{person}{Nicholas Kong}, \bibinfo{person}{Marti~A.
  Hearst}, {and} \bibinfo{person}{Maneesh Agrawala}.}
  \bibinfo{year}{2014}\natexlab{}.
\newblock \showarticletitle{Extracting References Between Text and Charts via
  Crowdsourcing}. In \bibinfo{booktitle}{\emph{Proceedings of the SIGCHI
  Conference on Human Factors in Computing Systems}} (Toronto, Ontario, Canada)
  \emph{(\bibinfo{series}{CHI '14})}. \bibinfo{publisher}{ACM},
  \bibinfo{address}{New York, NY, USA}, \bibinfo{pages}{31--40}.
\newblock
\showISBNx{978-1-4503-2473-1}
\urldef\tempurl%
\url{https://doi.org/10.1145/2556288.2557241}
\showDOI{\tempurl}


\bibitem[\protect\citeauthoryear{Large, Beheshti, Breuleux, and Renaud}{Large
  et~al\mbox{.}}{1995}]%
        {large:1995}
\bibfield{author}{\bibinfo{person}{Andrew Large}, \bibinfo{person}{Jamshid
  Beheshti}, \bibinfo{person}{Alain Breuleux}, {and} \bibinfo{person}{Andre
  Renaud}.} \bibinfo{year}{1995}\natexlab{}.
\newblock \showarticletitle{Multimedia and comprehension: The relationship
  among text, animation, and captions}.
\newblock \bibinfo{journal}{\emph{Journal of the American society for
  information science}} \bibinfo{volume}{46}, \bibinfo{number}{5}
  (\bibinfo{year}{1995}), \bibinfo{pages}{340--347}.
\newblock


\bibitem[\protect\citeauthoryear{Li, Jiang, and Shatkay}{Li
  et~al\mbox{.}}{2018}]%
        {li:2018}
\bibfield{author}{\bibinfo{person}{Pengyuan Li}, \bibinfo{person}{Xiangying
  Jiang}, {and} \bibinfo{person}{Hagit Shatkay}.}
  \bibinfo{year}{2018}\natexlab{}.
\newblock \showarticletitle{Extracting Figures and Captions from Scientific
  Publications}. In \bibinfo{booktitle}{\emph{Proceedings of the 27th ACM
  International Conference on Information and Knowledge Management}} (Torino,
  Italy) \emph{(\bibinfo{series}{CIKM '18})}. \bibinfo{publisher}{Association
  for Computing Machinery}, \bibinfo{address}{New York, NY, USA},
  \bibinfo{pages}{1595--1598}.
\newblock
\showISBNx{9781450360142}
\urldef\tempurl%
\url{https://doi.org/10.1145/3269206.3269265}
\showDOI{\tempurl}


\bibitem[\protect\citeauthoryear{{Liang} and {Huang}}{{Liang} and
  {Huang}}{2010}]%
        {5571352}
\bibfield{author}{\bibinfo{person}{J. {Liang}} {and} \bibinfo{person}{M.~L.
  {Huang}}.} \bibinfo{year}{2010}\natexlab{}.
\newblock \showarticletitle{Highlighting in Information Visualization: A
  Survey}. In \bibinfo{booktitle}{\emph{2010 14th International Conference
  Information Visualisation}}. \bibinfo{pages}{79--85}.
\newblock


\bibitem[\protect\citeauthoryear{Liu, Zhou, Pan, Qian, Cai, and Lian}{Liu
  et~al\mbox{.}}{2009}]%
        {shixia:2009}
\bibfield{author}{\bibinfo{person}{Shixia Liu}, \bibinfo{person}{Michelle~X.
  Zhou}, \bibinfo{person}{Shimei Pan}, \bibinfo{person}{Weihong Qian},
  \bibinfo{person}{Weijia Cai}, {and} \bibinfo{person}{Xiaoxiao Lian}.}
  \bibinfo{year}{2009}\natexlab{}.
\newblock \showarticletitle{Interactive, Topic-Based Visual Text Summarization
  and Analysis}. In \bibinfo{booktitle}{\emph{Proceedings of the 18th ACM
  Conference on Information and Knowledge Management}} (Hong Kong, China)
  \emph{(\bibinfo{series}{CIKM '09})}. \bibinfo{publisher}{Association for
  Computing Machinery}, \bibinfo{address}{New York, NY, USA},
  \bibinfo{pages}{543--552}.
\newblock
\showISBNx{9781605585123}
\urldef\tempurl%
\url{https://doi.org/10.1145/1645953.1646023}
\showDOI{\tempurl}


\bibitem[\protect\citeauthoryear{Liu, Zhou, Pan, Song, Qian, Cai, and Lian}{Liu
  et~al\mbox{.}}{2012}]%
        {shixia:2012}
\bibfield{author}{\bibinfo{person}{Shixia Liu}, \bibinfo{person}{Michelle~X.
  Zhou}, \bibinfo{person}{Shimei Pan}, \bibinfo{person}{Yangqiu Song},
  \bibinfo{person}{Weihong Qian}, \bibinfo{person}{Weijia Cai}, {and}
  \bibinfo{person}{Xiaoxiao Lian}.} \bibinfo{year}{2012}\natexlab{}.
\newblock \showarticletitle{TIARA: Interactive, Topic-Based Visual Text
  Summarization and Analysis}.
\newblock \bibinfo{journal}{\emph{ACM Trans. Intell. Syst. Technol.}}
  \bibinfo{volume}{3}, \bibinfo{number}{2}, Article \bibinfo{articleno}{25}
  (\bibinfo{date}{Feb.} \bibinfo{year}{2012}), \bibinfo{numpages}{28}~pages.
\newblock
\showISSN{2157-6904}
\urldef\tempurl%
\url{https://doi.org/10.1145/2089094.2089101}
\showDOI{\tempurl}


\bibitem[\protect\citeauthoryear{Matzen, Haass, Divis, Wang, and Wilson}{Matzen
  et~al\mbox{.}}{2017}]%
        {matzen2017data}
\bibfield{author}{\bibinfo{person}{Laura~E Matzen}, \bibinfo{person}{Michael~J
  Haass}, \bibinfo{person}{Kristin~M Divis}, \bibinfo{person}{Zhiyuan Wang},
  {and} \bibinfo{person}{Andrew~T Wilson}.} \bibinfo{year}{2017}\natexlab{}.
\newblock \showarticletitle{Data visualization saliency model: A tool for
  evaluating abstract data visualizations}.
\newblock \bibinfo{journal}{\emph{IEEE transactions on visualization and
  computer graphics}} \bibinfo{volume}{24}, \bibinfo{number}{1}
  (\bibinfo{year}{2017}), \bibinfo{pages}{563--573}.
\newblock


\bibitem[\protect\citeauthoryear{{M}icrosoft {Q} \& {A}}{{M}icrosoft {Q} \&
  {A}}{2020}]%
        {powerbi}
{M}icrosoft {Q} \& {A} \bibinfo{year}{2020}\natexlab{}.
\newblock \bibinfo{title}{{M}icrosoft {Q} \& {A}}.
\newblock
  \bibinfo{howpublished}{\url{https://powerbi.microsoft.com/en-us/documentation/powerbi-service-q-and-a/}}.
\newblock


\bibitem[\protect\citeauthoryear{Mittal, Roth, Moore, Mattis, and
  Carenini}{Mittal et~al\mbox{.}}{1995}]%
        {mittal:1995}
\bibfield{author}{\bibinfo{person}{Vibhu Mittal}, \bibinfo{person}{Steven
  Roth}, \bibinfo{person}{Johanna Moore}, \bibinfo{person}{Joe Mattis}, {and}
  \bibinfo{person}{Giuseppe Carenini}.} \bibinfo{year}{1995}\natexlab{}.
\newblock \showarticletitle{Generating Explanatory Captions for Information
  Graphics}.
\newblock \bibinfo{journal}{\emph{Proceeedings of the International Joint
  Conference on Artificial Intelligence}}, \bibinfo{pages}{1276--1283}.
\newblock


\bibitem[\protect\citeauthoryear{Nugent}{Nugent}{1983}]%
        {nugent:1983}
\bibfield{author}{\bibinfo{person}{Gwen~C Nugent}.}
  \bibinfo{year}{1983}\natexlab{}.
\newblock \showarticletitle{Deaf students' learning from captioned instruction:
  The relationship between the visual and caption display}.
\newblock \bibinfo{journal}{\emph{The Journal of Special Education}}
  \bibinfo{volume}{17}, \bibinfo{number}{2} (\bibinfo{year}{1983}),
  \bibinfo{pages}{227--234}.
\newblock


\bibitem[\protect\citeauthoryear{O'Brien and Lauer}{O'Brien and Lauer}{2018}]%
        {o2018testing}
\bibfield{author}{\bibinfo{person}{Shaun O'Brien} {and} \bibinfo{person}{Claire
  Lauer}.} \bibinfo{year}{2018}\natexlab{}.
\newblock \showarticletitle{Testing the susceptibility of users to deceptive
  data visualizations when paired with explanatory text}. In
  \bibinfo{booktitle}{\emph{Proceedings of the 36th ACM International
  Conference on the Design of Communication}}. \bibinfo{pages}{1--8}.
\newblock


\bibitem[\protect\citeauthoryear{{Ottley}, {Peck}, {Harrison}, {Afergan},
  {Ziemkiewicz}, {Taylor}, {Han}, and {Chang}}{{Ottley} et~al\mbox{.}}{2016}]%
        {ottley:2016}
\bibfield{author}{\bibinfo{person}{A. {Ottley}}, \bibinfo{person}{E.~M.
  {Peck}}, \bibinfo{person}{L.~T. {Harrison}}, \bibinfo{person}{D. {Afergan}},
  \bibinfo{person}{C. {Ziemkiewicz}}, \bibinfo{person}{H.~A. {Taylor}},
  \bibinfo{person}{P.~K.~J. {Han}}, {and} \bibinfo{person}{R. {Chang}}.}
  \bibinfo{year}{2016}\natexlab{}.
\newblock \showarticletitle{Improving Bayesian Reasoning: The Effects of
  Phrasing, Visualization, and Spatial Ability}.
\newblock \bibinfo{journal}{\emph{IEEE Transactions on Visualization and
  Computer Graphics}} \bibinfo{volume}{22}, \bibinfo{number}{1}
  (\bibinfo{year}{2016}), \bibinfo{pages}{529--538}.
\newblock


\bibitem[\protect\citeauthoryear{Pandey, Rall, Satterthwaite, Nov, and
  Bertini}{Pandey et~al\mbox{.}}{2015}]%
        {pandey2015deceptive}
\bibfield{author}{\bibinfo{person}{Anshul~Vikram Pandey},
  \bibinfo{person}{Katharina Rall}, \bibinfo{person}{Margaret~L Satterthwaite},
  \bibinfo{person}{Oded Nov}, {and} \bibinfo{person}{Enrico Bertini}.}
  \bibinfo{year}{2015}\natexlab{}.
\newblock \showarticletitle{How deceptive are deceptive visualizations? An
  empirical analysis of common distortion techniques}. In
  \bibinfo{booktitle}{\emph{Proceedings of the 33rd Annual ACM Conference on
  Human Factors in Computing Systems}}. \bibinfo{pages}{1469--1478}.
\newblock


\bibitem[\protect\citeauthoryear{{P}ew {R}esearch}{{P}ew {R}esearch}{2020}]%
        {pewresearch}
{P}ew {R}esearch \bibinfo{year}{2020}\natexlab{}.
\newblock \bibinfo{title}{{P}ew {R}esearch}.
\newblock \bibinfo{howpublished}{\url{https://www.pewresearch.org}}.
\newblock


\bibitem[\protect\citeauthoryear{Qian, Koh, Du, Kim, and Chan}{Qian
  et~al\mbox{.}}{2020}]%
        {qian2020formative}
\bibfield{author}{\bibinfo{person}{Xin Qian}, \bibinfo{person}{Eunyee Koh},
  \bibinfo{person}{Fan Du}, \bibinfo{person}{Sungchul Kim}, {and}
  \bibinfo{person}{Joel Chan}.} \bibinfo{year}{2020}\natexlab{}.
\newblock \showarticletitle{A Formative Study on Designing Accurate and Natural
  Figure Captioning Systems}. In \bibinfo{booktitle}{\emph{Extended Abstracts
  of the 2020 CHI Conference}}. \bibinfo{pages}{1--8}.
\newblock


\bibitem[\protect\citeauthoryear{{Srinivasan}, {Drucker}, {Endert}, and
  {Stasko}}{{Srinivasan} et~al\mbox{.}}{2019}]%
        {voder}
\bibfield{author}{\bibinfo{person}{A. {Srinivasan}}, \bibinfo{person}{S.~M.
  {Drucker}}, \bibinfo{person}{A. {Endert}}, {and} \bibinfo{person}{J.
  {Stasko}}.} \bibinfo{year}{2019}\natexlab{}.
\newblock \showarticletitle{Augmenting Visualizations with Interactive Data
  Facts to Facilitate Interpretation and Communication}.
\newblock \bibinfo{journal}{\emph{IEEE Transactions on Visualization and
  Computer Graphics}} \bibinfo{volume}{25}, \bibinfo{number}{1}
  (\bibinfo{year}{2019}), \bibinfo{pages}{672--681}.
\newblock
\urldef\tempurl%
\url{https://doi.org/10.1109/TVCG.2018.2865145}
\showDOI{\tempurl}


\bibitem[\protect\citeauthoryear{{T}ableau {P}ublic}{{T}ableau
  {P}ublic}{2020}]%
        {tableaupublic}
{T}ableau {P}ublic \bibinfo{year}{2020}\natexlab{}.
\newblock \bibinfo{title}{{T}ableau {P}ublic}.
\newblock \bibinfo{howpublished}{\url{https://public.tableau.com}}.
\newblock


\bibitem[\protect\citeauthoryear{{T}ableau {S}oftware}{{T}ableau
  {S}oftware}{2020}]%
        {tableau}
{T}ableau {S}oftware \bibinfo{year}{2020}\natexlab{}.
\newblock \bibinfo{title}{{T}ableau {S}oftware}.
\newblock \bibinfo{howpublished}{\url{http://www.tableau.com}}.
\newblock


\bibitem[\protect\citeauthoryear{Tufte}{Tufte}{1990}]%
        {tufte:1990}
\bibfield{author}{\bibinfo{person}{Edward Tufte}.}
  \bibinfo{year}{1990}\natexlab{}.
\newblock \bibinfo{booktitle}{\emph{Envisioning Information}}.
\newblock \bibinfo{publisher}{Graphics Press}, \bibinfo{address}{USA}.
\newblock
\showISBNx{0961392118}


\bibitem[\protect\citeauthoryear{Vartak, Madden, and Parmeswaran}{Vartak
  et~al\mbox{.}}{2015}]%
        {Vartak2015}
\bibfield{author}{\bibinfo{person}{Manasi Vartak}, \bibinfo{person}{Samuel
  Madden}, {and} \bibinfo{person}{Aditya~N Parmeswaran}.}
  \bibinfo{year}{2015}\natexlab{}.
\newblock \showarticletitle{{SEEDB : Supporting Visual Analytics with
  Data-Driven Recommendations}}.
\newblock  (\bibinfo{year}{2015}).
\newblock


\bibitem[\protect\citeauthoryear{{W}ashington {P}ost}{{W}ashington
  {P}ost}{2020}]%
        {washingtonpost}
{W}ashington {P}ost \bibinfo{year}{2020}\natexlab{}.
\newblock \bibinfo{title}{{W}ashington {P}ost}.
\newblock \bibinfo{howpublished}{\url{https://www.washingtonpost.com}}.
\newblock


\bibitem[\protect\citeauthoryear{Wei, Liu, Song, Pan, Zhou, Qian, Shi, Tan, and
  Zhang}{Wei et~al\mbox{.}}{2010}]%
        {wei:2010}
\bibfield{author}{\bibinfo{person}{Furu Wei}, \bibinfo{person}{Shixia Liu},
  \bibinfo{person}{Yangqiu Song}, \bibinfo{person}{Shimei Pan},
  \bibinfo{person}{Michelle~X. Zhou}, \bibinfo{person}{Weihong Qian},
  \bibinfo{person}{Lei Shi}, \bibinfo{person}{Li Tan}, {and}
  \bibinfo{person}{Qiang Zhang}.} \bibinfo{year}{2010}\natexlab{}.
\newblock \showarticletitle{TIARA: A Visual Exploratory Text Analytic System}.
  In \bibinfo{booktitle}{\emph{Proceedings of the 16th ACM SIGKDD International
  Conference on Knowledge Discovery and Data Mining}} (Washington, DC, USA)
  \emph{(\bibinfo{series}{KDD '10})}. \bibinfo{publisher}{Association for
  Computing Machinery}, \bibinfo{address}{New York, NY, USA},
  \bibinfo{pages}{153--162}.
\newblock
\showISBNx{9781450300551}
\urldef\tempurl%
\url{https://doi.org/10.1145/1835804.1835827}
\showDOI{\tempurl}


\bibitem[\protect\citeauthoryear{{W}ikipedia}{{W}ikipedia}{2020}]%
        {wikipedia}
{W}ikipedia \bibinfo{year}{2020}\natexlab{}.
\newblock \bibinfo{title}{{W}ikipedia}.
\newblock \bibinfo{howpublished}{\url{https://www.wikipedia.org}}.
\newblock


\bibitem[\protect\citeauthoryear{Wills and Wilkinson}{Wills and
  Wilkinson}{2010}]%
        {wills2010autovis}
\bibfield{author}{\bibinfo{person}{Graham Wills} {and} \bibinfo{person}{Leland
  Wilkinson}.} \bibinfo{year}{2010}\natexlab{}.
\newblock \showarticletitle{Autovis: automatic visualization}.
\newblock \bibinfo{journal}{\emph{Information Visualization}}
  \bibinfo{volume}{9}, \bibinfo{number}{1} (\bibinfo{year}{2010}),
  \bibinfo{pages}{47--69}.
\newblock


\bibitem[\protect\citeauthoryear{Xiong, van Weelden, and Franconeri}{Xiong
  et~al\mbox{.}}{2019}]%
        {xiong2019curse}
\bibfield{author}{\bibinfo{person}{Cindy Xiong}, \bibinfo{person}{Lisanne van
  Weelden}, {and} \bibinfo{person}{Steven Franconeri}.}
  \bibinfo{year}{2019}\natexlab{}.
\newblock \showarticletitle{The curse of knowledge in visual data
  communication}.
\newblock \bibinfo{journal}{\emph{IEEE Transactions on Visualization and
  Computer Graphics}} (\bibinfo{year}{2019}).
\newblock


\bibitem[\protect\citeauthoryear{Yu, Reiter, Hunter, and Sripada}{Yu
  et~al\mbox{.}}{2003}]%
        {yu:2003}
\bibfield{author}{\bibinfo{person}{Jin Yu}, \bibinfo{person}{Ehud Reiter},
  \bibinfo{person}{Jim Hunter}, {and} \bibinfo{person}{Somayajulu Sripada}.}
  \bibinfo{year}{2003}\natexlab{}.
\newblock \showarticletitle{SumTime-Turbine: A Knowledge-Based System to
  Communicate Gas Turbine Time-Series Data}. In
  \bibinfo{booktitle}{\emph{Developments in Applied Artificial Intelligence}},
  \bibfield{editor}{\bibinfo{person}{Paul W.~H. Chung}, \bibinfo{person}{Chris
  Hinde}, {and} \bibinfo{person}{Moonis Ali}} (Eds.).
  \bibinfo{publisher}{Springer Berlin Heidelberg}, \bibinfo{address}{Berlin,
  Heidelberg}, \bibinfo{pages}{379--384}.
\newblock
\showISBNx{978-3-540-45034-4}


\end{thebibliography}


\end{document}